\documentclass[journal]{IEEEtran}
\usepackage{amssymb}
\usepackage{amsmath}
\usepackage{algorithm,algorithmic}
\usepackage{cite}
\usepackage{graphicx}
\usepackage{blindtext}

\begin{document}
\title{Joint Radio Resource Allocation, 3D Placement and User Association of Aerial Base Stations in IoT Networks }
\author{Arman Azizi, Nader Mokari, Mohamad Reza Javan}

\maketitle

\begin{abstract}
In this paper, a novel method for joint radio resource allocation (RRA), three-dimensional placement (3DP), and user association of aerial base stations (ABSs) as a main problem in the internet of things (IoT) networks is proposed.
In our proposed model, we consider two schemes: a) line of sight (LoS) b) generalized. In the LoS scheme, all the ABSs should see the IoT users as LoS. In the generalized scheme, ABSs can see some of the IoT users as LoS and some of them as NLoS.  The main goal of this paper is to minimize the overal transmit power of the IoT users while satisfying some quality of service (QoS) constraints in uplink scenario. To solve the optimization problems and to convert the main problems with high complexity into the subproblems with lower complexity, we decompose them into two subproblems namely  3DP subproblem and joint RRA and user association (JRU) subproblem. The methods which we use to solve our proposed optimization problems are Semi Definite Relaxation (SDR) and Geometric Programming (GP). Finally, using simulations, we evaluate the performance of the proposed schemes for different values of the network parameters.
\end{abstract}
\vspace{-.6 cm}
\section{Introduction}\label{Introduction}

\subsection{State of the Art}
\subsubsection{Necessity to use UAVs}
The use of unmanned aerial vehicles (UAVs) in wireless communication networks
 has received remarkable attentions recently
\cite{abdulla2015toward,al2014modeling,al2014optimal,alzenad2016fso,bekmezci2013flying,bor2016efficient,chen2017caching,dawy2017toward,dhillon2017wide,fotouhi2016dynamic,han2009optimization,hayajneh2016optimal,kalantari2017backhaul,kosmerl2014base,li2015drone,li2017optimal,lien2011toward,merwaday2015uav,merwaday2016improved,motlagh2016low,mozaffari2015drone,mozaffari2016efficient,mozaffari2016mobile,mozaffari2016optimal,mozaffari2016unmanned,orfanus2016self,sharma2016uav,sharma2016uavs,soorki2016resource,wang2017improving,zhan2011wireless}.
%The most important challenge of UAV communications focuses on the air-to-ground (A2G) channel modeling. The LoS probability for A2G communication as a function of 3-D location of UAV compared to the two-dimensional (2-D) location of user was derived.
The use of UAVs can be useful in some regions  which have no cellular infrastructures or the areas in which building a cellular infrastructure is very expensive.
 The use of UAVs as aerial base stations (ABSs) in wireless communication networks makes lots of advantages.  ABSs can be deployed in higher altitudes compared to the terrestrial base stations. Therefore, they can see ground users with higher chance of line of sight (LoS) links. More over, ABSs can easily move to updated places, and therefore, they can be more flexible than  terrestrial base stations in the scenarios where users are mobile \cite{bekmezci2013flying}. The technical challenges in ABS communication networks can be classified into optimal placement, air to ground (A2G) channel modeling, resource management, energy efficiency, and  performance analysis \cite{al2014modeling,bekmezci2013flying,orfanus2016self}.
In the next generation of wireless communications, networks will require extra dense base stations deployments not only in 2-D area but also in 3-D space. Therefore, ABSs can play an important role for wireless cellular networks in overloaded cases. Furthermore, ABSs are more robust against environmental changes \cite{al2014modeling}.\\
\subsubsection{Necessity to use UAVs in IoT}
 IoT can change the ways of wireless communications with all of the devices. An IoT system uses intelligent interfaces for connecting IoT devices to each other at anytime and anywhere which use any network and any service \cite{motlagh2016low,ngu2017iot,angelakis2016allocation,razzaque2016middleware }. More over, UAVs are easy to deploy, capable of reprogramming during run time, capable of
 measuring anything anywhere and capable of having high mobility, therefore, they can
 be chosen to provide many applications, such as service delivery,
 farming, pollution mitigation, and rescue operations \cite{motlagh2016low, yin2017offline,motlagh2017uav }. According to the above definitions, UAVs are the efficient option to choose for IoT network, and hence, they can play an important role.  Furthermore, at the same time, they can provide extra services when they
 are equipped with some special devices (e.g., cameras, 
 actuators and sensors)
  \cite{motlagh2016low,yin2017offline,motlagh2017uav}.
In the other hand, the transmit power of IoT devices are low in comparison with other traditional networks, and hence, IoT devices can not be able to communicate in long range. More over, ABSs can update their locations due to the new locations of IoT devices. Therefore, ABSs can collect IoT data from the IoT devices and transmit it to other devices which are out of transmission range. Accordingly, ABSs can play an important role in IoT networks which have battery-limited devices \cite{dawy2017toward,lien2011toward,dhillon2017wide}.
\subsection{Related Works}
The Related works to this paper can be classified into the following items:
\subsubsection{Placement}
%The work in \cite{mozaffari2016unmanned} studies the scenario including UAV and device-to-device (D2D) communication considering coverage and rate.
%The A2G channel model is presented in \cite{al2014modeling},  \cite{feng2006path} and \cite{holis2008elevation} for UAV communications.
The authors in \cite{bor2016efficient} maximize the number of users which covered by a single ABS  by finding the efficient 3D placement for the ABS.
In \cite{al2014optimal}, the authors obtain the optimal altitude deployment of a UAV in order to maximize the coverage.
In \cite{kosmerl2014base}, the authors enhance the coverage performance in public safety communications by obtaining the optimal deplyment of UAVs. 
%In \cite{al2014optimal}, an analytical approach to determine the coverage-optimal UAV altitude has been developed for a single-UAV system.
In \cite{hayajneh2016optimal}, the authors use the sigmoid function LoS model to optimize the UAV height in different performance metrics.
 In \cite{alzenad2016fso}, the authors design the ABS enabled heterogeneous networks (HetNets) by  3D placements of ABSs  as one of the important factors.  
 In \cite{mozaffari2015drone}, the optimal altitude of a single UAV is found to achieve a required coverage with minimum transmit power.
In \cite{mozaffari2016efficient},the authors find  the optimal coverage range and hovering altitude of UAVs to minimize the transmit power of them.
In \cite{sharma2016uavs}, the authors propose the next generation
heterogeneous network including cooperative UAVs. Accordingly, they find the optimal placement and  optimal distribution of UAVs to optimize the overall
network delays.
In \cite{sharma2016uav}, the authors study on a model which uses density and cost
functions to compute the areas with higher demands, and hence, the UAVs
are deployed based on these cost functions.
In \cite{merwaday2016improved}, the locations of
ABSs are  optimized in order to maximize the network throughput over a given geographical area.
In \cite{merwaday2015uav}, the authors maximize the fifth percentile
capacity of the network by optimizing the
locations of the UAVs.
\\
\subsubsection{Power Allocation}
In \cite{wang2017improving}, the authors investigate a secure wireless network based on physical layer security technique considering UAV as a relay. In order to  maximize the secrecy rate, they  optimize the transmit power of the source and the UAV as a relay  while satisfying a sequence
of information-causality constraints. These
constraints ensure that the relay cannot forward
the undecoded data.
\subsubsection{Joint Power Allocation and Spectrum Allocation}
In \cite{soorki2016resource}, the authors present a scenario including cluster heads-UAVs
communications in M2M networks. They propose an efficient
scheduling and resource allocation mechanism in order to minimize the transmit power of cluster heads while
satisfying rate requirements of M2M devices.
In \cite{li2017optimal}, the authors investigate a cellular network
with multi-layer UAVs. In order to minimize the packet transmission delay, the resources allocation mechanism proposed.
\subsubsection{Joint Placement and User Association} 
In \cite{han2009optimization}, the authors study the optimal deployment of UAVs and association of  static ground users to UAVs in order to meet the users’ rate requirements.
The work in \cite{mozaffari2016optimal} investigates a communication network based on  multiple UAVs in downlink transmissions considering UAV efficient deployment and user association.
%In \cite{mozaffari2016mobile} the authors apply multi-objective optimization to UAV networks, using the sigmoid LoS model to characterize the received signal strength. Note that in the above works the UAV locations are assumed to either be known a priori or are found as part of an optimization problem. 
In \cite{fotouhi2016dynamic}, the authors study on the UAV networks using the sigmoid LoS model in order to characterize the received signal strength. Accordingly, they propose the multi-objective optimization. Furthermore, the UAVs locations are found as a part of an optimization problem or they are assumed to be known. In \cite{kalantari2017backhaul},
 the authors present a more comprehensive placement and user association problem of a single UAV-BS. In \cite{mozaffari2016mobile}, the authors present an IoT network based on UAVs. They study on 
 deploying the UAVs efficiently in order to minimize the power consumption of the
ground IoT devices by considering
the required bit error rate. Furthermore, they present the efficient
association of the ground IoT devices to UAVs.
In \cite{chen2017caching}, the authors
maximize the users' quality of experience (QoE) while minimizing the transmit power of the UAVs.
\subsubsection{Joint Placement and Spectrum Allocation}
In \cite{li2015drone}, the authors propose a  
multi-hop D2D network based on UAV in order 
 to develop the coverage of network. They show that deploying the UAV in an efficient way  increases the data rate assuming the transmit power or distance is beyond the threshold.

\subsubsection{Adaptive Modulation}
In \cite{zhan2011wireless}, the authors investigate a scenario in which UAVs can be used
as relays between ground devices and a ground base station. They propose a method in order to control the heading angle of UAVs considering space-time coding and adaptive modulation, and hence, they  optimize the performance of the ground-to-relay links. In \cite{abdulla2015toward}, the authors maximize the energy efficiency %(throughputper energy)
 in the network in which ground nodes are capable of adaptive modulation. Furthermore, they show how mobility pattern of UAVs can affect on adaptive modulation. 
% For the mobility pattern intrinsic to the UAVs, they  show how adaptive modulation is affected. 

%\subsubsection{IoT}
%%UAVs in the latter category are in particular drawing the attention of the internet of things (IoT) community: the authors of \cite{motlagh2016low} suggest that the majority of UAVs in IoT applications will be miniature devices that operate at heights below 300 meters.
%\\ %The reason for this is that miniature, low altitude UAVs offer lower cost, more flexible deployment and they make use of airspace which is far less utilised by manned aircraft and is therefore subject to more relaxed regulations \cite{atkins2010risk}.

 %where a number of constraints including maximum PL, backhaul data rate and UAV bandwidth limit were considered.% However \cite{kalantari2017backhaul} solve the respective optimization problems using exhaustive search, which is not practically applicable. 
%In \cite{bor2016new}, a novel framework of multi-tier drone-BSs complementing terrestrial heterogeneous networks (HetNets) is envisioned, and advancements and challenges related to the operation and management of drone-BSs are discussed.
% In \cite{sathyanarayanan2016designing}, design and implementation challenges of an aerial network of base stations is reported and the capabilities of different aerial platforms for carrying wireless communication systems is reviewed.
 
\subsection{Our Contribution}
The main contribution of this paper is to develop a novel
approach for inteligently deploying multiple ABSs while minimizing the overall ABS
transmit power needed to satisfy the users’ data rate in uplink scenario. Furthermore, subcarrier allocation and adaptive modulation are used in this scenario. %and derive,
%jointly, the locations
%of ABSs that minimize the required transmit power. \\
Our contribution of this paper can be listed as follows:
\begin{itemize}
		\item Joint RRA, 3DP, and User Association: 
	The main factor of ABSs which makes them useful  is the ability to move, and hence, the location of ABSs is not fixed. Therefore, we should see what is the impact of finding the location of ABSs on RRA and user association. Accordingly, it is necessary to find the radio variables, placement variables, and user association variables in a joint method. 
	To the best of our knowledge, there is no work  which consider the joint RRA, 3DP, and user association.
	\item Comparing A2G channel models: 
	To the best of our knowledge, there is no work which provide the comparison between two A2G channel models namely only the LoS scheme and the generalized scheme. Finally, we know which one is better for our IoT Scenario.  
	%\item Holding LoS probability as a function of elevation angle and altitude of ABS in problem formulations
	\item Multi ABS, multi IoT users, multi subcarrier system, and adaptive modulation: 
	To the best of our knowledge, there is no work which has all the above items together. In this paper,
 we show the impact of different modulation orders on overal transmit power of IoT users. Furtheremore, we show the impact of ABS numbers on the overal transmit power of the IoT users and  the average ABS altitude.

\item Converting the non-convex problems into convex with relaxation methods:
We propose two main problems for our scenario which both of them are non-convex and intractable, and hence, we convert them into convex problems with semidefinite relaxation (SDR) and geometric programming (GP).

\item Computational Complexity: We obtain the computational complexity of our proposed schemes.

\end{itemize}

\section{System Model}\label{system model}
Consider an IoT system consisting of a set $\mathcal{I} = \{ 1,...,I\}$
including $I$ IoT users deployed within a geographical area. In
this system, a set $\mathcal{J} =\{ 1,...,J\}$ including $J$ ABSs should be
deployed to collect the data from the ground devices in the
uplink. Furthermore, a set $\mathcal{M} =\{ 1,...,M\}$ including $M$ modulation orders should be assigned to each user for uplink transmition. The $m^{th}$ modulation order shows that we use  $2^{m+1}$PSK for transmission. The locations of user
$i $ and ABS $j$
are, respectively, given by $({\hat x_i} ;{\hat y_i}  )$ and $(x_{j} ; y_{j} ; h_j)$. We assume that devices transmit in the uplink using
orthogonal frequency division multiple access (OFDMA).
Note
that, we consider a network in which the locations
of devices are known to a control center such as
a central cloud server. The ground IoT devices can be mobile
(e.g., smart cars) and their data availability can be intermittent
(e.g., sensors).
\begin{figure}[!h]
\centering
\includegraphics[width=9cm,height=8cm]{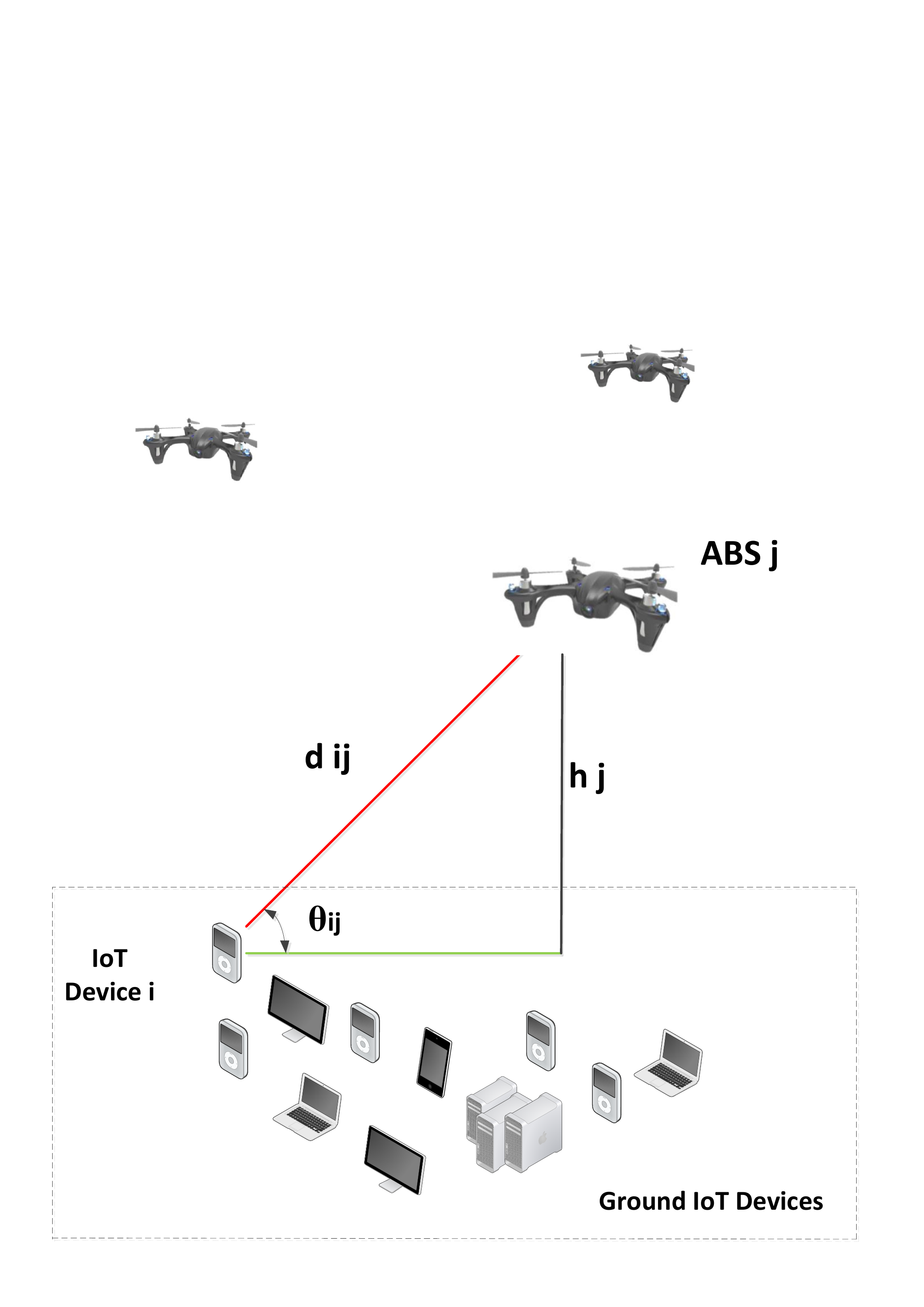}
\caption{System Model}\label{figmodel}
\end{figure}
 For A2G communications, each device can typically have a LoS
view towards a specific ABS with a given probability. This
LoS probability depends on the environment, location of the
device and the ABS, and the elevation angle between the
device and the ABS \cite{al2014modeling}. One suitable expression for the LoS
probability between ABS $j$ and user $i$ is given by \cite{al2014modeling}:
\begin{equation}\label{plos}
\begin{array}{*{20}{c}}
{{{Pr}_{{\text{LoS}},ij}} = \frac{1}{{1 + \alpha \exp ( - \beta (\frac{{180}}{\pi }{\theta _{ij}} - \alpha ))}}},&{\forall i,j},
\end{array}
\end{equation}
where $\alpha$ and $\beta$ are constant values which depend on the carrier frequency and type of environment such as rural, urban, or
dense urban, and $\theta_{ij}$ is the elevation angle that is defined as follows:  
\begin{equation}\label{theta}
\begin{array}{*{20}{c}}
{{\theta _{ij}} = \frac{{180}}{\pi } \times {{\sin }^{ - 1}}(\frac{{{h_j}}}{{{d_{ij}}}})},&{\forall i,j},
\end{array}
\end{equation}
and $d_{ij}$ is the distance between device $i$ and ABS $j$ given by
\begin{equation}\label{d}
\begin{array}{*{20}{c}}
{{d_{ij}} = \sqrt {{{({x_j} - {\hat x_i}  )}^2} + {{({y_j} - {\hat y_i}  )}^2} + h_j^2} },&{\forall i,j}.
\end{array}
\end{equation}

The path loss expressions for LoS and NLoS connections are as follows \cite{al2014modeling}:
\begin{equation}\label{Llos}
\begin{array}{*{20}{c}}
{{L_{\text{{LoS}},ij}} = 10n\log (\frac{{4\pi {f_c}{d_{ij}}}}{c}) + {\xi _\text{{LoS}}}},&{\forall i,j},
\end{array}
\end{equation}
\begin{equation}\label{Lnlos}
\begin{array}{*{20}{c}}
{{L_{\text{{NLoS}},ij}} = 10n\log (\frac{{4\pi {f_c}{d_{ij}}}}{c}) + {\xi _\text{{NLoS}}}},&{\forall i,j},
\end{array}
\end{equation}
where $n$ is the path loss exponent, $L_{\text{{LoS}},ij}$ and $L_{\text{{NLoS}},ij}$ are the average path loss for LoS and
NLoS links, respectively. ${\xi _\text{{LoS}}}$ and ${\xi _\text{{NLoS}}}$ are the average additional loss in addition to the free space propagation loss which depend on the environment, $c$ is the speed of light, and $f_c$ is the carrier frequency.
Finally, the average path loss as a function of the ABS altitude
and coverage radius can be written as \cite{al2014optimal}:
\begin{equation}\label{average pathloss}
\overline L (x_j,y_j,h_j) = {{Pr}_{\text{{LoS}},ij}}{L_{\text{{LoS}},ij}} + {{Pr}_{\text{{NLoS}},ij}}{L_{\text{{NLoS}}.ij}},
\end{equation}
where ${Pr}_{\text{{NLoS}},ij}$ is NLoS probability between ABS $j$ and user $i$.
Therefore, we obtain a closed-form for average path loss by substituting \eqref{plos}, \eqref{Llos}, and \eqref{Lnlos} into \eqref{average pathloss} as follows:
\begin{equation}\label{practical pathloss}
\hspace{-0.3cm}\bar L({x_j},{y_j},{h_j}) =10n\log (\frac{{4\pi {f_c}{d_{ij}}}}{c}) + {{Pr}_{\text{{LoS}},ij}}({\xi _\text{{LoS}}} - {\xi _\text{{NLoS}}}) + {\xi _\text{{NLoS}}}.
\end{equation}

For minimum phase shift keying (MPSK) modulation, the bit error rate expression is given by \cite{goldsmith2005wireless}:
\begin{equation}\label{BER}
{BER_{imj}} = \frac{2}{{m+1}}Q(\sqrt {\frac{{2P_{imj}^{\text{receive}}}}{{{\hat{r}_{ij}}{N_0}}}} \sin (\frac{\pi }{2^{m+1}})),
\end{equation}
where $P^{\text{receive}}_{imj}$ is the received power from the $j^{th}$ ABS to  the $i^{th}$ user when using  the $m^{th}$ modulation order for transmission, $\hat{r}_{ij}$ is the symbol rate of the $i^{th}$ user to the $j^{th}$ ABS, $N_0$ is the noise power spectral density and $Q(.)$ is the Q-function. In the next step, we propose the closed-form formulation for transmit power of the following schemes. 
\begin{table}\label{table2}
	\centering
	\caption{Simulation Parameters  }
	\label{table-1}
	\begin{tabular}{|c|c|c|c|}
		\hline
		
		$i$ & IoT user indicator\\
		\hline

		$j$ & ABS indicator\\
		
		\hline

		$m$& Modulation order indicator  \\
		
		%	\hline
		%	$r_{imj}$& Average symbol rate& 200 kbps\\
		
		\hline
		$l$& Subcarrier indicator \\
		\hline
		$({\hat x_i} ;{\hat y_i}  )$&2D location of the $i^{th}$ user \\
		\hline
		$(x_{j} ; y_{j} ; h_j)$& 3D location of the $j^{th}$ ABS \\
		\hline
		$\alpha$& Constant value depends on environment\\
		\hline
		$\beta$& Constant value depends on environment\\
		\hline
		$\theta_{ij}$& Elevation angle of ABS\\
		\hline
		$Pr_{{\text{LoS}},ij}$& LoS probability between ABS $j$ and user $i$ \\
		\hline
		$d_{ij}$& Distance between device $i$ and ABS $j$ \\
		\hline
		$n$& Path loss exponent\\
		\hline
		$L_{\text{{LoS}},ij}$& Average path loss for LoS link \\
		\hline
			$L_{\text{{NLoS}},ij}$& Average path loss for NLoS link \\
		\hline
			${\xi _\text{{LoS}}}$& Average additional loss of LoS link \\
		\hline
			${\xi _\text{{NLoS}}}$& Average additional loss of NLoS link \\
		\hline
				$c$& Speed of light  \\
		\hline
			$f_c$& Carrier frequency \\
		\hline
			${\overline L(x_j,y_j,h_j)}$& Average Path loss \\
		\hline
			$Pr_{\text{{NLoS}},ij}$& NLoS probability between ABS $j$ and user $i$ \\
		\hline
			$P^{\rm recieve}_{imj}$& Received power from the $j^{th}$ ABS \\
		\hline
			 $\hat{r}_{ij}$& Symbol rate of the $j^{th}$ UAV to  the $i^{th}$ user \\
		\hline
			$P_{imj}^{\rm transmit}$& Transmit power of IoT user\\
		\hline
			${r_{imj}} $&  Transmission bit rate \\
		\hline
			$N_0$ & Noise power spectral density \\
		\hline
			$\varepsilon$& Threshold for LoS probability \\
		\hline
			$\rho _{imj}^l$&Binary variable for RRA and user association  \\
		\hline
				$\tau_i $& Threshold of transmission rate for each user \\
		\hline
			$\delta$& Bit error rate requirement \\
		\hline
	\end{tabular}
\end{table}

\subsection{The Case with only the LoS}
By using \eqref{Llos}, \eqref{BER}, $
P_{imj}^{\text{transmit}} = P_{imj}^{\text{receive}} \times {10^{\frac{{{L_{LoS,ij}}}}{{10}}}}$, and  considering $n=2$, the minimum transmit power of ABS $j$
needed to reach a bit error rate requirement of
$\delta$ is given by
\begin{equation}\label{objective0}
P_{imj}^{\rm {transmit} } = {A_m} \cdot {r_{imj}} \cdot d_{ij}^2,
\end{equation}
where $P^{\text{transmit}}_{imj}$ is the transmit power and  $
{r_{imj}} = {\hat{r}_{ij}}(m + 1)$ is the transmission bit rate from the the $i^{th}$ user to the $j^{th}$ ABS by using the $m^{th}$ modulation order for transmission, respectively.
 $A_m$ is a constant that only changes by the modulation order as follows:
 \begin{multline}
{A_m} = {({Q^{ - 1}}(\frac{1}{2}\delta (m + 1)) \times \frac{1}{{\sin (\frac{\pi }{{{2^{m + 1}}}})}})^2} \times \frac{1}{{(m + 1)}} \\ \times \frac{1}{2}{N_0} \times {(\frac{{4\pi {f_c}}}{c})^2} \times {10^{\frac{{{\xi _{}}LoS}}{{10}}}},
 \end{multline}
where $Q^{-1}(.)$ is the inverse Q-function.

In the LoS model, the necessary condition for connecting a device to the UAV is to have a LoS probability greater than a threshold ($\varepsilon$ is closed to 1). In other words, $Pr_{LoS}(\theta_{ij})\geq\varepsilon$, and hence, $\theta_{ij}\geq Pr_{LoS}^{-1}(\varepsilon)$ leading to:
\begin{equation}\label{LoS}
\begin{array}{*{20}{c}}
{{d_{ij}} \le \frac{{{h_j}}}{{\sin (Pr_{LoS}^{ - 1}(\varepsilon ))}}},&{\forall i,j}.
\end{array}
\end{equation}
Note that \eqref{LoS} guarantees the user $i$ can connect to ABS $j$ if the distance between them is not greater than 
$\frac{{{h_j}}}{{\sin (Pr_{LoS}^{ - 1}(\varepsilon ))}}$.
Our goal is to maximizing the overal transmit power of  the IoT users in the joint scenario including RRA, 3DP and user association. Accordingly, the objective function is given by:
\begin{equation}\label{objective1}
\mathop {\min }\limits_{{\boldsymbol{\rho }},{\bf{x}},{\bf{y}},{\bf{h}}} \sum\limits_{j = 1}^J {\sum\limits_{m = 1}^M {\sum\limits_{i = 1}^I {\sum\limits_{l = 1}^L {P_{imj}^{\text{transmit}}} } } } \rho _{imj}^l,
\end{equation}
where $\rho _{imj}^l$ is a binary variable that is one if there is a connection between the $i^{th}$ user and the $j^{th}$ UAV using the $m^{th}$ modulation order in the $l^{th}$ subcarrier, and is zero otherwise. By substituting \eqref{objective0} into \eqref{objective1}, our optimization problem can be formulated as:
\begin{subequations}\label{eq7:power_Problem_formulation} 
	\begin{align}
	\mathop {\min }\limits_{{\boldsymbol{\rho }},{\bf{x}},{\bf{y}},{\bf{h}}} \sum\limits_{j = 1}^J {\sum\limits_{m = 1}^M {\sum\limits_{i = 1}^I {\sum\limits_{l = 1}^L {{A_m}{r_{imj}}} } } } d_{ij}^2\rho _{imj}^l
	\end{align}
	\vspace{-0.5cm}
	\begin{align}\label{binaryrho}
\hspace{-2.8cm}\begin{array}{*{20}{c}}
{\begin{array}{*{20}{c}}
	{s.t.}&{\rho _{imj}^l \in \{ 0,1\} }
	\end{array}},&{\forall i,m,j,l},
\end{array}
	\end{align}
		\vspace{-0.5cm}
	\begin{align}\label{sumrate}
	\hspace{-1.5cm}\begin{array}{*{20}{c}}
	{\sum\limits_{j = 1}^J {\sum\limits_{m = 1}^M {\sum\limits_{l = 1}^L {{r_{imj}}} } } \rho _{imj}^l \ge \tau_i },&{\forall i},
	\end{array}
	\end{align}
		\vspace{-0.5cm}
		\begin{align}\label{independentuav}
		\hspace{0.6cm}
	\begin{array}{*{20}{c}}
		{\rho _{imj}^l+\rho _{im'j'}^{l'} \le 1},&{\forall i,j \ne j',m,l,m',l'},
		\end{array}
	\end{align}
		\vspace{-0.5cm}
		\begin{align}\label{independentuav0}
		\hspace{-1.2cm}
\begin{array}{*{20}{c}}
		{\rho _{imj}^l+\rho _{im'j}^l \le 1},&{\forall i,j,l,m \ne m'},
		\end{array}
	\end{align}
		\vspace{-0.5cm}
		\begin{align}\label{independentsub}
		\hspace{-2.5cm}
		{\begin{array}{*{20}{c}}
			{\sum\limits_{m = 1}^M {\sum\limits_{i = 1}^I {\sum\limits_{j = 1}^J {\rho _{imj}^l = 1} } } },&{\forall l},
			\end{array}}
	\end{align}
		\vspace{-0.5cm}
		\begin{align}\label{LoS1}
				\hspace{-1.5cm}
	\begin{array}{*{20}{c}}
	{\rho _{imj}^l{d_{ij}} \le \frac{{{h_j}}}{{\sin (Pr_{LoS}^{ - 1}(\varepsilon ))}},}&{\forall i,j,m,l.}
	\end{array}
	\end{align}
	\end{subequations}
Constraint \eqref{sumrate} states that the transmission rate for each user is greater than or equal to a threshold $\tau_i$. Constraint \eqref{independentuav} guarantees that only one UAV can be assigned to each user. Constraint \eqref{independentuav0}  shows that for each ABS,  each user and each subcarrier, we can use at most one modulation order. Constraint \eqref{independentsub} shows that the subcarriers are orthogonal in OFDMA. More over, it guarantees that for each subcarrier we have at most one modulation and one assignment of UAV to user.  Constraint \eqref{LoS1} is the necessary condition for connecting a device to a UAV to have a LoS probability greater than a threshold.

\subsection{Generalized Scheme (Jonit NLoS and LoS)}
In this model, by using \eqref{practical pathloss}, \eqref{BER},  $P_{imj}^{\text{transmit}} = P_{imj}^{\text{receive}} \times {10^{\frac{{\bar L({x_j},{y_j},{h_j})}}{{10}}}}$, and considering $n=2$, the minimum transmit power of UAV $j$
needed to reach a bit error rate requirement of
$\delta$ is given by:
\begin{equation}\label{practical power}
P_{imj}^{\text{transmit}} = {B_m} \times {r_{imj}} \times d_{ij}^2 \times {10^{\frac{{{Pr_{LoS,ij}}({\xi _{LoS}} - {\xi _{NLoS}})}}{{10}}}},
\end{equation}
where $B_m$ is a constant that only changes by the modulation order as follows:
\begin{multline}
{B_m} = {({Q^{ - 1}}(\frac{1}{2}\delta (m + 1)) \times \frac{1}{{\sin (\frac{\pi }{{{2^{m + 1}}}})}})^2} \times \frac{1}{{(m + 1)}} \times \frac{1}{2}{N_0}\\ \times {(\frac{{4\pi {f_c}}}{c})^2}{\rm{ }} \times {10^{\frac{{{\xi _{NLoS}}}}{{10}}}}.
\end{multline}
Therefore, our optimization problem in this case can be demonstrated as follows:
\begin{subequations}
	\begin{multline}
	\mathop {\min }\limits_{{\bf{\rho }},{\bf{x}},{\bf{y}},{\bf{h}}} \sum\limits_{j = 1}^J {\sum\limits_{m = 1}^M {\sum\limits_{i = 1}^I {\sum\limits_{l = 1}^L {{B_m}{r_{imj}}\rho _{imj}^l} } } } d_{ij}^2\\\times{10^{\frac{\eta }{{1 + \alpha \exp ( - \beta (\frac{{180}}{\pi }{{\sin }^{ - 1}}(\frac{{{h_j}}}{{{d_{ij}}}}) - \alpha ))}}}}
	\end{multline}
	\begin{align}
	\begin{array}{*{20}{c}}
	\hspace{-4.4cm}{s.t.}&{\eqref{binaryrho}-\eqref{independentsub},}
	\end{array}
	\end{align}
\end{subequations}
where $\eta  = \frac{{({\xi _{LoS}} - {\xi _{NLoS}})}}{{10}}$.
\section{SOLUTION}\label{solution}
\subsection{The Case with only the Line of Sight}
Our optimization problem is Mixed Integer Nonlinear Programming (MINLP), nonconvex, and NP-hard. To solve the resulting optimization problem, it is decomposed into 3DP subproblem and JRU subproblem which are iteratively solved until convergence to a local solution. 

 \subsubsection{3DP}
 The 3DP subproblem can be formulated as Quadratically Constrained Quadratic Program (QCQP)  for a fixed vector $\boldsymbol{\rho}$ as follows:
 \begin{subequations}\label{sub3dp}
 	\begin{align}\label{3dp}
 	\hspace{-0.5cm}
 	\mathop {\min }\limits_{\bf{x},\bf{y},\bf{h}} \sum\limits_{j = 1}^J {\sum\limits_{m = 1}^M {\sum\limits_{i = 1}^I {\sum\limits_{l = 1}^L {{A_m}{r_{imj}}} } } } \rho _{imj}^ld_{ij}^2
 	\end{align}
 	\begin{align}
\begin{array}{*{20}{c}}
{s.t.}&{\begin{array}{*{20}{c}}
	{\rho _{imj}^l{d_{ij}} \le \frac{{{h_j}}}{{\sin (Pr_{LoS}^{ - 1}(\varepsilon ))}},}&{\forall i,j \in \Omega ,\forall m,l,}
	\end{array}}
\end{array}
 	\end{align}
 \end{subequations}
where $
\begin{array}{*{20}{c}}
{\Omega  = \{ (i,j)}
\end{array}|\begin{array}{*{20}{c}}
{\rho _{imj}^l \in {\boldsymbol{\rho _0}}}&{\forall m,l}
\end{array}\} 
$ and $\boldsymbol{\rho_{0}}$ is an initial matrix  for iterative method.

Note that by using \eqref{d}, \eqref{sub3dp} can be transformed into \eqref{subgpobj}  as follows:  
\begin{subequations}\label{subgpobj}
	\begin{multline}\label{gpobj}
	\hspace{-0.1cm}
	\mathop {\min }\limits_{{\mathbf{x}},{\mathbf{y}},{\mathbf{h}}} \sum\limits_{j = 1}^J {\sum\limits_{m = 1}^M {\sum\limits_{i = 1}^I {\sum\limits_{l = 1}^L {{A_m}{r_{imj}}} } } } \rho _{imj}^l({({x_j} - {{\hat x}_i})^2} + {({y_j} - {{\hat y}_i})^2}\\ + h_j^2)
	\end{multline}
\begin{align}\label{qcqpcon}
\hspace{-0.5cm}
\begin{array}{*{20}{c}}
{s.t.}&{\begin{array}{*{20}{c}}
	{\rho _{imj}^l{{({x_j} - {{\hat x}_i})}^2} + \rho _{imj}^l{{({y_j} - {{\hat y}_i})}^2} + \rho _{imj}^lh_j^2(1 - }\\
	{\frac{1}{{{{\sin }^2}(Pr_{LoS}^{ - 1}(\varepsilon ))}}) \le 0}.
	\end{array}}
\end{array}
\end{align}
\end{subequations}
As we can see from \eqref{subgpobj}, the optimization problem is a QCQP whose general form is given by\cite{boyd2004convex,luo2010semidefinite}:\\
\begin{subequations}\label{nonconvexqcqp}
		\begin{multline}
	 \mathop {\min }\limits_{\boldsymbol{v}} \frac{1}{2}{\boldsymbol{v^T}}{\boldsymbol{W_0}}\boldsymbol{v} + \boldsymbol{Q_0^T}\boldsymbol{v} + {\tilde{r}_0}~~~~~~~~~~~~~~
	\end{multline}
		\begin{multline}
s.t.~~\frac{1}{2}\boldsymbol{{v^T}}\boldsymbol{{W_i}}\boldsymbol{v} + \boldsymbol{Q_i^T}\boldsymbol{v} + {\tilde{r}_i}\leq0,~~ \forall i.~~~~~~~~~~~~
	\end{multline}
\end{subequations}
Given \eqref{subgpobj}, we have:
\begin{equation}
\boldsymbol{v} = \left[ {\begin{array}{*{20}{c}}
	{{x_1}}&{{y_1}}&{{h_1}}&.&{\begin{array}{*{20}{c}}
		.&{{x_J}}
		\end{array}}&{{y_J}}&{{h_J}}
	\end{array}} \right]_{3J \times 1}^T,
\end{equation}
\begin{equation}
{{\mathbf{W}}_{\mathbf{0}}} = {\left[ {\begin{array}{*{20}{c}}
  {{\boldsymbol{\Gamma} _{\mathbf{1}}}}&0&0&0 \\ 
  0&.&0&0 \\ 
  0&0&.&0 \\ 
  0&0&0&{{\boldsymbol{\Gamma} _{\mathbf{J}}}} 
\end{array}} \right]_{{\mathbf{3J}} \times {\mathbf{3J}}}},
\end{equation}
where $\boldsymbol{\Gamma_{j}}$ is a matrix which can be written as follows:
\begin{equation}
\boldsymbol{\Gamma_{j}} = \left[ {\begin{array}{*{20}{c}}
	{2LMI{\omega _j}}&0&0\\
	0&{2LMI{\omega _j}}&0\\
	0&0&{2LMI{\omega _j}}
	\end{array}} \right]_{3\times 3},~~\forall j\in J, 
\end{equation}
and ${\omega _j} = \sum\limits_i {\sum\limits_m {\sum\limits_l {{A_m}{r_{imj}}\rho _{imj}^l} } }$. Also we have:
\begin{equation}
{W_i} = {\left[ {\begin{array}{*{20}{c}}
  {{\Upsilon _{i1}}}&0&0&0 \\ 
  0&.&0&0 \\ 
  0&0&.&0 \\ 
  0&0&0&{{\Upsilon _{iJ}}} 
\end{array}} \right]_{3J \times 3J}},
\end{equation}
where ${\Upsilon _{ij}}$ is a matrix that can be written as follows:
\begin{equation}
{\Upsilon _{ij}} = \left[ {\begin{array}{*{20}{c}}
  {2LM{\vartheta _{ij}}}&0&0 \\ 
  0&{2LM{\vartheta _{ij}}}&0 \\ 
  0&0&{2LM\kappa{\vartheta _{ij}} } 
\end{array}} \right]_{3J\times 3J},
\end{equation}
where $
\kappa  = 1 - \frac{1}{{{{\sin }^2}(Pr_{LoS}^{ - 1}(\varepsilon ))}}
$ and ${\vartheta _{ij}} = \sum\limits_m {\sum\limits_l {{A_m}{r_{imj}}} } \rho _{imj}^l$ ,\\
Furthermore, we have:
\begin{equation}
\boldsymbol{{Q_0}} = {[\begin{array}{*{20}{c}}
  \boldsymbol{{{\Lambda _1}}}&.&.&\boldsymbol{{{\Lambda _J}} }
\end{array}]^{T}_{3J \times 1}}~ ,
\end{equation} 
where $\boldsymbol{\Lambda_j}$ is a matrix that can be formulated as follows:\\\\
$
\boldsymbol{{\Lambda _j}} = {\left[ {\begin{array}{*{20}{c}}
  { - 2LM\sum\limits_i {{\vartheta _{ij}}{{\hat x}_i}} }&{ - 2LM\sum\limits_i {{\vartheta _{ij}}{{\hat y}_i}} }&0 
\end{array}} \right]_{_{1 \times 3}}}$,
\\

$
{{\mathbf{Q}}_{\mathbf{i}}} = [\begin{array}{*{20}{c}}
  {{\Theta _{i1}}}&.&.&{{\Theta _{iJ}}} 
\end{array}]_{3J \times 1 }^T
$,\\\\$
\boldsymbol{{\Theta _{ij}}} = {\left[ {\begin{array}{*{20}{c}}
  { - 2LM{\vartheta _{ij}}{{\hat x}_i}}&{ - 2LM{\vartheta _{ij}}{{\hat y}_i}}&0 
\end{array}} \right]_{1 \times 3}}$,
\\
$
{\tilde{r}_0} = LMJ\sum\limits_i {{\gamma _i}({{\hat x}^2}_i + {{\hat y}^2}_i)}  $,\\
$
{\gamma _i} = \sum\limits_m {\sum\limits_j {\sum\limits_l {{A_m}{r_{imj}}\rho _{imj}^l} } } $, and
$
{\tilde{r}_i} = LMJ({{\hat x}^2}_i + {{\hat y}^2}_i){\gamma _i}
$.\\
Note that $\boldsymbol{W_i}$ is not a positive semidefinite matrix, and hence, the QCQP problem in \eqref{nonconvexqcqp} is nonconvex.
Therefore, in order to solve the NP-hard problem \eqref{nonconvexqcqp}, we should convert the nonconvex QCQP problem into a semidefinite programming problem by semidefinite programming relaxation (SDR) method.

First of all, we should convert the non-homogeneous QCQP problem into homogeneous QCQP problem. The homogeneous form of problem \eqref{nonconvexqcqp} can be formulated as follows \cite{luo2010semidefinite}:
\begin{subequations}\label{homo}
\begin{equation}
\begin{array}{*{20}{c}}
{\mathop {\min }\limits_{{\boldsymbol{v}},{ a}} }
\end{array}{\frac{1}{2}} \left[ {\begin{array}{*{20}{c}}
	{{\boldsymbol{v^T}}}&a 
	\end{array}} \right]\left[ {\begin{array}{*{20}{c}}
	{\boldsymbol{{W_0}}}&{\boldsymbol{{Q_0}}} \\ 
	{\boldsymbol{Q_0^T}}&0 
	\end{array}} \right]{\left[ {\begin{array}{*{20}{c}}
		{\boldsymbol{{v^T}}}&a 
		\end{array}} \right]^T} + {\tilde{r}_0}
\end{equation}
\begin{equation}
\begin{array}{*{20}{c}}
{s.t.}
\end{array}{\frac{1}{2}} \left[ {\begin{array}{*{20}{c}}
	{\boldsymbol{{v^T}}}&a
	\end{array}} \right]\left[ {\begin{array}{*{20}{c}}
	{\boldsymbol{{W_i}}}&{\boldsymbol{{Q_i}}} \\ 
	\boldsymbol{{Q_i^T}}&0 
	\end{array}} \right]{\left[ {\begin{array}{*{20}{c}}
		{\boldsymbol{{v^T}}}&a
		\end{array}} \right]^T} + {\tilde{r}_i},
\end{equation}
\begin{equation}
\frac{1}{2}\left[ {\begin{array}{*{20}{c}}
	{\boldsymbol{{v^T}}}&a
	\end{array}} \right]\left[ {\begin{array}{*{20}{c}}
	0&0\\
	0&1
	\end{array}} \right]{\left[ {\begin{array}{*{20}{c}}
		{\boldsymbol{{v^T}}}&a
		\end{array}} \right]^T} + {\tilde{r}_i}.
\end{equation}
\end{subequations}
The problem \eqref{homo} can be formulated as the following equivalent problem:
\begin{subequations}\label{uU}
	\begin{equation}
\begin{array}{*{20}{c}}
{\mathop {\min }\limits_{\boldsymbol{{u},{U}}} }
\end{array}{\frac{1}{2}}tr(\boldsymbol{{T_0}{U}}) + {\tilde{r}_0}
	\end{equation}
	\begin{equation}
\begin{array}{*{20}{c}}
	{s.t.}&{{\frac{1}{2}}tr(\boldsymbol{{T_i}{U}}) + {\tilde{r}_i} \le 0},
	\end{array}
	\end{equation}
	\begin{equation}\label{uuT}
\boldsymbol{U} = \boldsymbol{u^T}\boldsymbol{{u}},
	\end{equation}
	\begin{equation}\label{a2=1}
\boldsymbol{u}\left[ {\begin{array}{*{20}{c}}
		0&0\\
		0&1
		\end{array}} \right]\boldsymbol{{u^T} }= 1,
	\end{equation}
\end{subequations}
where $
\boldsymbol{u} = {\left[ {\begin{array}{*{20}{c}}
		{\boldsymbol{{v^T}}}&a
		\end{array}} \right]_{1 \times (1 + 3J)}}$ ,
$\boldsymbol{{T_0}} = \left[ {\begin{array}{*{20}{c}}
	{{{\bf{W}}_{\bf{0}}}}&{{{\bf{Q}}_{\bf{0}}}}\\
	{{\bf{Q}}_{\bf{0}}^{\bf{T}}}&0
	\end{array}} \right]$ and
$\boldsymbol{{T_i}} = \left[ {\begin{array}{*{20}{c}}
{{{\bf{W}}_{\bf{i}}}}&{{{\bf{Q}}_{\bf{i}}}}\\
{{\bf{Q}}_{\bf{i}}^{\bf{T}}}&0
\end{array}} \right]$. The constraint \eqref{uuT} is equivalent to\\
${\rm {rank}}(\boldsymbol{U}) = 1$ and $
\boldsymbol{U} \succeq 0$ (shows the matrix $\boldsymbol{U}$ is semi definite positive (SDP)). Furthermore, the constraint \eqref{a2=1} is equivalent to $a^2=1$.

In the next step, we should convert the  nonconvex homogeneous QCQP problem \eqref{uU} into an SDP problem. The constraint ${\rm rank}(\boldsymbol{U}) = 1$ makes the optimization problem \eqref{uU} nonconvex. Therefore, we should relax the optimization problem \eqref{uU} by ignoring ${\rm {rank} }(\boldsymbol{U}) = 1$. Finally, the convex relaxed problem can be written as follows:
\begin{subequations}\label{relaxed}
	\begin{equation}
\begin{array}{*{20}{c}}
	{\mathop {\min }\limits_{\boldsymbol{U }}}
	\end{array}{\frac{1}{2}}tr(\boldsymbol{{T_0}U}) + {\tilde{r}_0}
	\end{equation}
		\begin{equation}
\begin{array}{*{20}{c}}
	{s.t.}&\frac{1}{2}{tr(\boldsymbol{{T_i}U}) + {\tilde{r}_i} \le 0},
	\end{array}
	\end{equation}
	\begin{equation}
\boldsymbol{U} \ge 0,
	\end{equation}
	\begin{equation}\label{a2==1}
	tr(\boldsymbol{HU}) = 1,
	\end{equation}
\end{subequations}
where the constraint \eqref{a2==1} is equivalent to $a^2=1$ because 
${\rm \boldsymbol {H}} = {\left[ {\begin{array}{*{20}{c}}
		0&0\\
		0&1
		\end{array}} \right]_{(3J + 1) \times (3J + 1)}}$. By solving the problem \eqref{relaxed},\vspace{0.1cm} we reach to optimal $\boldsymbol{U}$ denoted $\boldsymbol{U^*}$. Then, we shoud find $\boldsymbol{u^*}$ by Gaussian Randomization Procedure (GRP)\cite{luo2010semidefinite}.
\begin{algorithm}\label{alg}
	\caption{Gaussian Randomization Procedure (GRP)}
	{\textbf{s1:}}~ Choose a feasible solution $\boldsymbol{U^*}$ for the relaxed homogeneous QCQP problem \eqref{relaxed}.
	\\\\{\textbf{s2:}}~For $g=1,...,G$ generate $\boldsymbol{k_g}\sim \rm \mathcal{N}(0,\boldsymbol{U^*})$ with zero mean and nonzero covariance.
	\\\\{\textbf{s3:}}~Determine $g^*$ in 
	${\rm {argmin} }\left\{ {{\boldsymbol{k}}_{\boldsymbol{{g}}}^{\boldsymbol{{T}}}{{\boldsymbol{{T}}}_{\boldsymbol{{0}}}}{{\boldsymbol{{k}}}_{\boldsymbol{{g}}}} + {{{\tilde{r}}}_{{0}}}} \right\}$ for the homogeneous QCQP.
 	\\\\ {\textbf{s4:}}~Output $\boldsymbol{u^*=k_g^*}$ for the homogeneous QCQP.
\end{algorithm}\\
As we know that 
$\boldsymbol{u} = \left[ {\begin{array}{*{20}{c}}
	{\boldsymbol{{v^T}}}&a
	\end{array}} \right]$, we can find $\boldsymbol{v^*}$. Therefore, the relaxed convex-3DP problem can be solved by some existing optimization tools such as CVX.

\subsubsection{Jonit RRA and User Association (JRU)}
The JRU subproblem can be formulated as Binary Linear Programming (BLP) for fixed $\boldsymbol{x,y,h}$ as follows:
\begin{subequations}
	\begin{align}\label{obj2}
\mathop {\min }\limits_{\boldsymbol{\rho}}  \sum\limits_{j = 1}^J {\sum\limits_{m = 1}^M {\sum\limits_{i = 1}^I {\sum\limits_{l = 1}^L {{A_m}{r_{imj}}} } } } \rho _{imj}^ld_{ij}^2
	\end{align}
	\begin{align}
\hspace{-3.5cm}{s.t.}~~~\eqref{binaryrho}-\eqref{LoS1}.
	\end{align}
\end{subequations}
Therefore, JRU subproblem as a Binary Linear Programming can be solved by some existing optimization tools such as NOMAD.
Finally, the main problem can be solved by Algorithm 2. 
\begin{algorithm}\label{alg}
	\caption{Iterative procedure of obtaining optimal solution}
	{\textbf{s1:}}~ Initialize $\boldsymbol{\rho}$ from the feasible set.
	\\\\{\textbf{s2:}}~ Calculate $\boldsymbol{x^{*}}$,$\boldsymbol{y^{*}}$,$\boldsymbol{h^{*}}$ from 3DP subproblem for fixed $\boldsymbol{\rho}$.
	\\\\{\textbf{s3:}}~Calculate obj1 and obj2 from \eqref{3dp} and \eqref{obj2}, respectively.
	\\\\ {\textbf{s4:}}~\textbf{for} $t=1$ to $T$ {\textbf{do}}
	\\\\ {\textbf{s4:}}~ \textbf{While} $\left| {obj1-obj2} \right|\ge\sigma$ or $t < T$ calculate $\boldsymbol{\rho^{*}}$ from the JRU subproblem for fixed $\boldsymbol{x,y,h}$ and  $\boldsymbol{x^{*}}$,$\boldsymbol{y^{*}}$,$\boldsymbol{h^{*}}$ from 3DP subproblem for fixed $\boldsymbol{\rho}$.
		\\\\ {\textbf{s4:}}~~~Set $t=t+1$
			\\\\ {\textbf{s4:}}~{\textbf{end for}}
	\\\\ {\textbf{s5:}}~ ($\boldsymbol{\rho^{*}}$,$\boldsymbol{x^{*}}$,$\boldsymbol{y^{*}}$,$\boldsymbol{h^{*}}$) is the optimal solution.
	% \label{alg2}
\end{algorithm}
where $t$ is the iteration number, $T$ is the upper bound for the iteration number, $\sigma$ is a number close to zero,
 obj1 and obj2 are the objectives of 3D placement subproblem and joint RRA and user association subproblem, respectively.
\subsection{Generalized Model}
Our optimization problem in the general case is also a MINP which is nonconvex and NP-hard. Therefore, to solve the problem, we decompose it into 3DP subproblem and  JRU subproblem. 
\subsubsection{3DP}
The 3DP subproblem can be formulated as geometric programming problem \cite{boyd2007tutorial,boyd2004convex} for a fixed $\boldsymbol{\rho}$ as follows:
\begin{multline}\label{25}
\mathop {\min }\limits_{{\mathbf{x}},{\mathbf{y}},{\mathbf{h}}} \sum\limits_{j = 1}^J {\sum\limits_{m = 1}^M {\sum\limits_{i = 1}^I {\sum\limits_{l = 1}^L {{B_m}{r_{imj}}} } } } \rho _{imj}^ld_{ij}^2\\\times{10^{\frac{\eta }{{1 + \alpha \exp ( - \beta (\frac{{180}}{\pi }{{\sin }^{ - 1}}(\frac{{{h_j}}}{{{d_{ij}}}}) - \alpha ))}}}}
\end{multline}
 The 3DP subproblem can be transformed into the following problem:
\begin{subequations}
\begin{multline}\label{26a}
\mathop {\min }\limits_{{\bf{x}},{\bf{y}},{\bf{h}},{{\bf{t}}_{\bf{0}}},{{\bf{t}}_{\bf{1}}},{{\bf{f}}_{\bf{0}}}} \sum\limits_{j = 1}^J {\sum\limits_{m = 1}^M {\sum\limits_{i = 1}^I {\sum\limits_{l = 1}^L {} } } } {B_m}{r_{imj}}\rho _{imj}^l(t_{0,ij}^2 + t_{1,ij}^2 + h_j^2)\\\times{10^{\frac{\eta }{{1 + \alpha \exp ( - \beta (\frac{{180}}{\pi }{{\sin }^{ - 1}}(\frac{{{h_j}}}{{{d_{ij}}}}) - \alpha ))}}}}
\end{multline}
   \begin{align}\label{taghrib}
\hspace{-4cm}
s.t.\begin{array}{*{20}{c}}
{\frac{1}{2}{x_j}({\hat x_i} ){^{ - \frac{1}{2}}}t_{0,ij}^{ - \frac{1}{2}} = 1},&{\forall i,j\in \Omega},
\end{array}
	\end{align}
	\begin{align}\label{taghrib0}
\hspace{-3.7cm}\begin{array}{*{20}{c}}
{\frac{1}{2}{y_j}({\hat y_i}){^{ - \frac{1}{2}}}t_{1,ij}^{ - \frac{1}{2}} = 1},&{\forall i,j\in \Omega},
\end{array}
	\end{align}
%		\begin{align}
%	\begin{array}{*{20}{c}}
%	\hspace{-4cm}{s.t.}&{\eqref{taghrib},\eqref{taghrib0}},
%	\end{array}
%	\end{align}
	\begin{align}\label{f0}
\begin{array}{*{20}{c}}
{f_{0,ij}^{ - 2}t_{0,ij}^2h_j^{ - 2} + f_{0,ij}^{ - 2}t_{1,ij}^2h_j^{ - 2} + f_{0,ij}^{ - 2} \leqslant 1},&{\forall i,j\in \Omega},
\end{array}
	\end{align}
\end{subequations}
where $d_{ij}^2$ in \eqref{25} is transformed into $t_{0,ij}^2 + t_{1,ij}^2 + h_j^2$ in \eqref{26a} as expalined before in the LoS scheme. \eqref{f0} is the posynomial form of $\frac{{{d_{ij}}}}{{{h_j}}}\leqslant f_{0,ij}$.\\ 

%One may question whether or not the term $
%{\sin ^{ - 1}}(\frac{1}{{{f_{0,ij}}}})$  in \eqref{26a} can be approximated as monomial or posynomial. This Question is answered with Proposition 1.

\textbf{\textit{Proposition 1}:} $f$ can be approximated by a monomial if and only if $F(y) = \log\begin{array}{*{20}{c}}{f({e^y})}\end{array}$can be approximated by an affine function. Furthuremore, $f$ can be approximated by a generalized posynomial if and only if $F$ can be approximated by a convex function \cite{boyd2007tutorial}.\\ 
By Proposition 1, we can show that the function $
{\sin ^{ - 1}}(\frac{1}{{{f_{0,ij}}}})$  can be approximated by a monomial if the function $
\log ({\sin ^{ - 1}}(\exp ( - {f_{0,ij}})))
$ is affine. It can be shown that this function is affine for ${f_{0,ij}} \geqslant 0.2$ and this condition always is established because $
1 \leqslant \frac{{{d_{ij}}}}{{{h_j}}} \leqslant {f_{0,ij}}$. Therefore, we have $
{\sin ^{ - 1}}(\frac{1}{{{f_{0,ij}}}}) \approx {\mu _2}f_{0,ij}^{{\omega _2}}$.\\
By substituting  $
{\sin ^{ - 1}}(\frac{1}{{{f_{0,ij}}}}) \approx {\mu _2}f_{0,ij}^{{\omega _2}}$ into \eqref{26a} and using the following approximations \cite{boyd2007tutorial}:
\begin{equation}\label{28}
\exp (\frac{{180}}{\pi }\beta {\mu _2}f_{0,ij}^{{\omega _2}}) \approx {(1 + \frac{{\frac{{180}}{\pi }\beta {\mu _2}f_{0,ij}^{{\omega _2}}}}{\psi })^\psi },
\end{equation}
\begin{equation}\label{29}
{10^{^{^{\eta {\mu _3}f_{1,ij}^{\psi  - {\omega _3}}}}}} \approx {(1 + \frac{{\eta {\mu _3}f_{1,ij}^{\psi  - {\omega _3}}\log _e^{10}}}{\phi })^\phi },
\end{equation}
\begin{equation}\label{30}
f_{1,ij}^\psi  + \alpha \exp (\alpha \beta ) \approx {\mu _3}f_{1,ij}^{{\omega _3}},
\end{equation}
where \eqref{28} and \eqref{29} are approximated for large amount of $\psi$ and $\phi$, respectively, our 3DP subproblem can be demonstrated as a GP problem. 
Furthermore, \eqref{30} shows monomial approximation for $f_{1,ij}^\psi  + \alpha \exp (\alpha \beta )$ . Finaly, the 3DP subproblem in the form of GP can be obtained as follows:
\begin{subequations}
	 \begin{multline}
	\hspace{-0.3cm}
\mathop {\min }\limits_{\scriptstyle{\bf{x}},{\bf{y}},{\bf{h}},{{\bf{t}}_{\bf{0}}},{{\bf{t}}_{\bf{1}}},\hfill\atop
	\scriptstyle{{\bf{f}}_{\bf{0}}},{{\bf{f}}_{\bf{1}}},{{\bf{f}}_{\bf{2}}}\hfill} \sum\limits_{j = 1}^J {\sum\limits_{m = 1}^M {\sum\limits_{i = 1}^I {\sum\limits_{l = 1}^L {{B_m}{r_{imj}}\rho _{imj}^l(t_{0,ij}^2 + t_{1,ij}^2 + h_j^2)} } } }\\\times f_{2,ij}^\phi
	\end{multline}
	\begin{align}
\begin{array}{*{20}{c}}
\hspace{-4.5cm}{s.t.}&{\eqref{taghrib},\eqref{taghrib0},\eqref{f0}},
\end{array}
\end{align}
	\begin{align}\label{f2}
	\hspace{-1.7cm}\begin{array}{*{20}{c}}
	{f_{1,ij}^{ - 1} + \frac{{180}}{\pi }\beta {\mu _2}f_{0,ij}^{{\omega _2}}{\psi ^{ - 1}}f_{1,ij}^{ - 1} \leqslant 1},&{\forall i,j\in \Omega} ,
	\end{array}
	\end{align}
	\begin{align}\label{f1}
	\hspace{-1.8cm}
	\begin{array}{*{20}{c}}
	{f_{2,ij}^{ - 1} + \frac{{\eta {\mu _3}f_{1,ij}^{\psi  - {\omega _3}}\log _e^{10}}}{\phi }f_{2,ij}^{ - 1} \leqslant 1},&{\forall i,j\in \Omega} ,
	\end{array}
	\end{align}
\end{subequations}
where \eqref{f2} and \eqref{f1} come from the exponential terms that approximated for large amount of $\psi$ and $\phi$ in \eqref{28} and \eqref{29}, respectively.
Therefore, the 3DP subproblem as a geometric programming can be solved by some existing optimization tools such as CVX.

 \subsubsection{JRU} The JRU subproblem can be formulated as Binary Linear Programming (BLP) for fixed $\boldsymbol{x,y,h}$ as follows:
 \begin{subequations}
\begin{multline}
\mathop {\min }\limits_{{\boldsymbol{\rho}}} \sum\limits_{j = 1}^J {\sum\limits_{m = 1}^M {\sum\limits_{i = 1}^I {\sum\limits_{l = 1}^L {{A_m}{r_{imj}}} } } } \rho _{imj}^ld_{ij}^2\\\times{10^{\frac{\eta }{{1 + \alpha \exp ( - \beta (\frac{{180}}{\pi }{{\sin }^{ - 1}}(\frac{{{h_j}}}{{{d_{ij}}}}) - \alpha ))}}}}
\end{multline}
\begin{align}
\begin{array}{*{20}{c}}
\hspace{-4.5cm}{s.t.}&{\eqref{binaryrho}-\eqref{independentsub}},
\end{array}
\end{align}
 \end{subequations}
Therefore, the JRU subproblem as a BLP can be solved by some existing optimization tools such as NOMAD.
Finally, the main problem can be solved by an iterative algorithm similar to Algorithm 2.
\section{Computational Complexity}
Here, we discuss the computational complexity of our proposed optimization problems, namely, LoS scheme and generalized scheme. Moreover, each of the proposed methods is decomposed into two subproblems namely JRU and 3DP. Therefore, we should compute the complexity of each of the subproblems in both of the proposed methods.

In the 3DP subproblem of LoS scheme, we use SDR in order to relax the 3DP subproblem, and hence, its computational complexity can be formulated as follows:
\begin{equation}
\max {\left\{ {3J + 1,I + 1} \right\}^3}{(3J + 1)^{0.5}}\log ({\raise0.7ex\hbox{$1$} \!\mathord{\left/
		{\vphantom {1 {\tilde \mu }}}\right.\kern-\nulldelimiterspace}
	\!\lower0.7ex\hbox{${\tilde \mu }$}}),
\end{equation}
where $\tilde \mu > 0$ is used for the given accuracy solution of interior point
method (IPM). SDR is a computationally efficient approximation approach
to QCQP in the sense that its complexity is polynomial in
the problem size and the number of constraints.
The computational complexity of JRU subproblem in the LoS scheme can be written as follows: 
\begin{equation}\label{comp-jru-los}
\frac{{\log (\frac{{4IMJL + I + L}}{{\tilde t\tilde \mu }})}}{{\log (\tilde \xi )}},
\end{equation}
where $\tilde \mu$ is used for the accuracy updating of interior point
method (IPM), $\tilde \xi$ is the stopping criterion for IPM, and $\tilde t$
is the initial point for approximating the accuracy of IPM.

The computational complexity formulation of JRU subproblem in the generalized scheme is the same as \eqref{comp-jru-los}. Furthermore, the computational complexity of 3DP subproblem in the generalized scheme can be written as follows:
\begin{equation}
\frac{{\log (\frac{{5IJ}}{{\tilde t\tilde \mu }})}}{{\log (\tilde \xi )}}.
\end{equation}
The computational complexity of the proposed problems can be listed in the Table \ref{table-2} as follows:
\begin{table}[h]\label{table2}
	\centering
	\caption{{Computational Complexity } }
	\label{table-2}
	\begin{tabular}{|c|c|c|c|}
		\hline
		
		\textbf{Kind of Problem}& \textbf{Computational Complexity}  \\
		\hline

		3DP Subproblem of LoS scheme &$
		\begin{array}{*{20}{c}}
		{\max {{\left\{ {3J + 1,I + 1} \right\}}^3} \times }\\
		{{{(3J + 1)}^{0.5}}\log (\frac{1}{{\tilde \mu }})}
		\end{array}$\\
		
		\hline

		JRU Subproblem of LoS Scheme & $
		\begin{array}{*{20}{c}}
		{}\\
		{\frac{{\log (\frac{{4IMJL + I + L}}{{\tilde t\tilde \mu }})}}{{\log (\tilde \xi )}}}
		\end{array}$  \\
		
		%	\hline
		%	$r_{imj}$& Average symbol rate& 200 kbps\\
		
		\hline
		3DP Subproblem of Generalized Scheme&$\frac{{\log (\frac{{5IJ}}{{\tilde t\tilde \mu }})}}{{\log (\tilde \xi )}}$ \\
		\hline
		JRU Subproblem of Generalized Scheme& $
		\begin{array}{*{20}{c}}
		{}\\
		{\frac{{\log (\frac{{4IMJL + I + L}}{{\tilde t\tilde \mu }})}}{{\log (\tilde \xi )}}}
		\end{array}$ \\
		\hline
	\end{tabular}
\end{table}
\section{Simulation Results}\label{simulation results}
In our simulations, the IoT users are deployed in an area of size  
$\rm {1~ km} \times \rm{1 ~km} $. We consider this scenario in an urban environment with $\alpha=9.61$ and $\beta=0.16$ at 2.1 GHz carrier frequency.
Table \ref{table-1} defines the simulation parameters. Note that we reach to all the results after a large number of independent runs.
\begin{table}[h]\label{table2}
	\centering
	\caption{Simulation Parameters  }
	\label{table-1}
	\begin{tabular}{|c|c|c|c|}
		\hline
		
		$f_c$ & carrier frequency & 2.1GHz\\
		\hline

		$\delta$ &bit error rate&$10^{-8}$\\
		
		\hline

		$N_0$&noise power spectral density & -170 dBm \\
		
		%	\hline
		%	$r_{imj}$& Average symbol rate& 200 kbps\\
		
		\hline
		$\xi _{\text{{LoS}}}$&aditional loss for LoS &1.6dB\\
		\hline
		$\xi _{\text{{NLoS}}}$&aditional loss for NLoS &23dB\\
		\hline
		$n$& path loss exponent& 2\\
		\hline
		$\alpha$& constant value for $P_{LoS}$& 9.61\\
		\hline
		$\beta$& constant value for $P_{LoS}$& 0.16\\
		\hline
	\end{tabular}
\end{table} 

Fig. 2 indicates that which user is connected to which ABS and shows the efficient locations of ABSs should  be deployed. In this figure, we have 5 ABSs to support 80 IoT users.
Note that, each ABS can be allocated to at most
25 IoT users because of resource block limitation. Therefore, the maximum size of the IoT user clusters is 25. It is obvious that the
location of IoT users affects on the number of
users per cluster and also the efficient locations of the ABSs. Note that, the minimum and maximum IoT user cluster sizes are 10
and 20, respectively. Fig. 3 shows the 3-D view of ABSs deployment, and hence, we can see that ABSs altitude affects on the number of IoT users to support. 

Fig. 4 shows the average ABS altitude versus the number of the ABSs for two different channel models which we use in our system model as LoS scheme and generalized scheme. Note that, by increasing the number of ABSs, the average ABS altitude decreases. By increasing the 
 number of ABSs, overlapping between the coverage
 regions of the ABSs increases. Therefor, the coverage radius of ABSs must be decreased by reducing their height. More over, in the generalized scheme the ABSs can see the users in dense area as LoS and see other users as NloS. Therefore, the average ABS altitude reduces in generalized scheme in comparison with the LoS scheme. Due to the fact that ABSs in higher altitude need more transmit power, and hence, the overall transmit power of IoT users in  generalized scheme is lower than the overall transmit power in LoS scheme as shown in Fig. 5.
   
Fig. 5 shows the overal transmit power of IoT users versus
the number of ABSs.
In this figure, the performance of the proposed methods is
compared with the fixed ABSs case which considers that the locations of ABSs are known. In the fixed ABSs case, assuming a uniform distribution of
IoT users, we fix the location of ABSs at an altitude of 550 m,
and then, we assign each IoT user to the nearest ABS. Therefore, in the fixed ABSs case we have only RRA optimization problem. In the next step, we can see that the maximum size
of each cluster decreases as the number of ABSs increases. Furthermore, we can see that the generalized scheme can be more efficient than the LoS scheme because the average altitude of ABSs decreases in the generalized scheme in comparison with the LoS method. Note that, the proposed generalized model does not spoil the LoS ability of ABSs. For example, consider a number of users that most of them are densified in an area but a few numbers are far from the majority. If we use the LoS scheme for deployment of ABSs, we must deploy them in high altitudes that they can see all the users with LoS, But we can waiver the minority and deploy the ABSs in a way that they can only see the majority of IoT users as LoS. Not only this assumption does not spoil the LoS ability of ABSs, but also it can  make our method more efficient in practical cases. As expected, increasing the number of ABSs
reduces the overal transmit power of IoT users.

Fig. 6 shows the overal transmit power of the IoT users versus the number of the ABSs. We can see, using two modulation order (QPSK+8PSK) is more efficient in comparison with (QPSK). Using different modulation orders makes the ABSs more flexible than before in order to choose the best option for connection, and hence, the Overal transmit power of  the IoT users decreases in comparison with the case using one modulation order. As expected, the total transmit power of the generalized model is lower than the LoS method.

\begin{figure}[!h]\label{fig2}
\hspace{-1.3cm}
\includegraphics[scale=0.29]{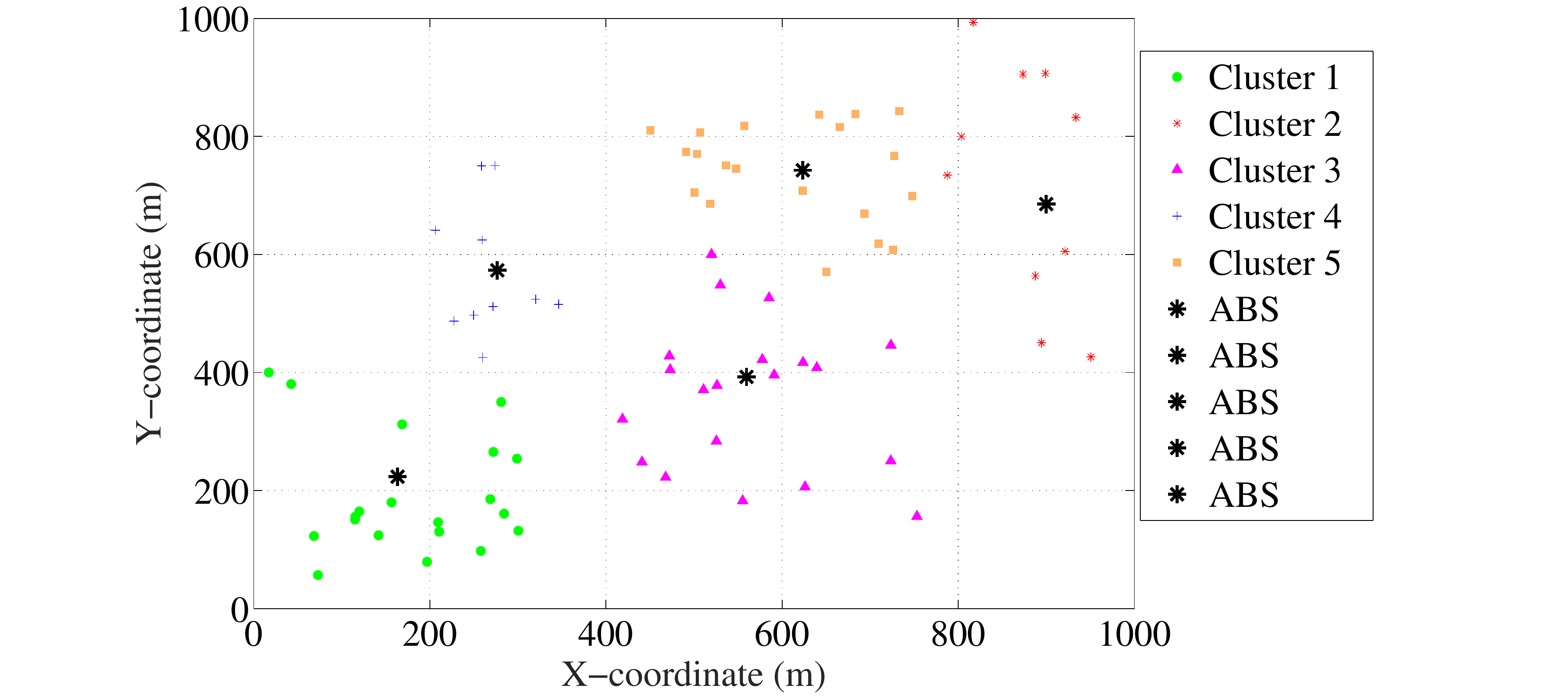}
\caption{User association and 2D placement of ABSs considering the case with only the LoS }\label{}
\end{figure}

\begin{figure}[!h]
\hspace{-1.5cm}
\includegraphics[scale=0.33]{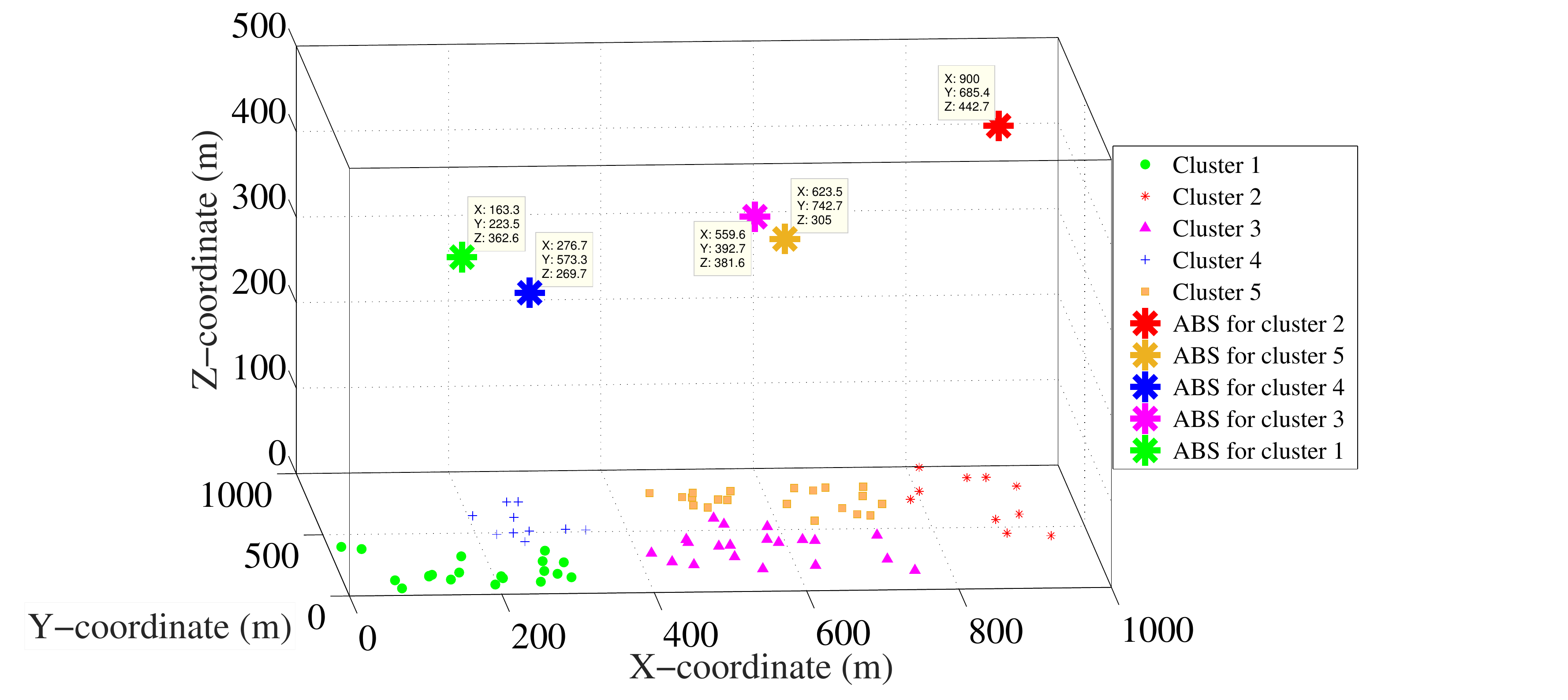}
\caption{User association and 3D placement of ABSs considering the case with only the LoS}\label{}
\end{figure}

\begin{figure}[!h] 
\hspace{0.2cm}
\includegraphics[scale=0.32]{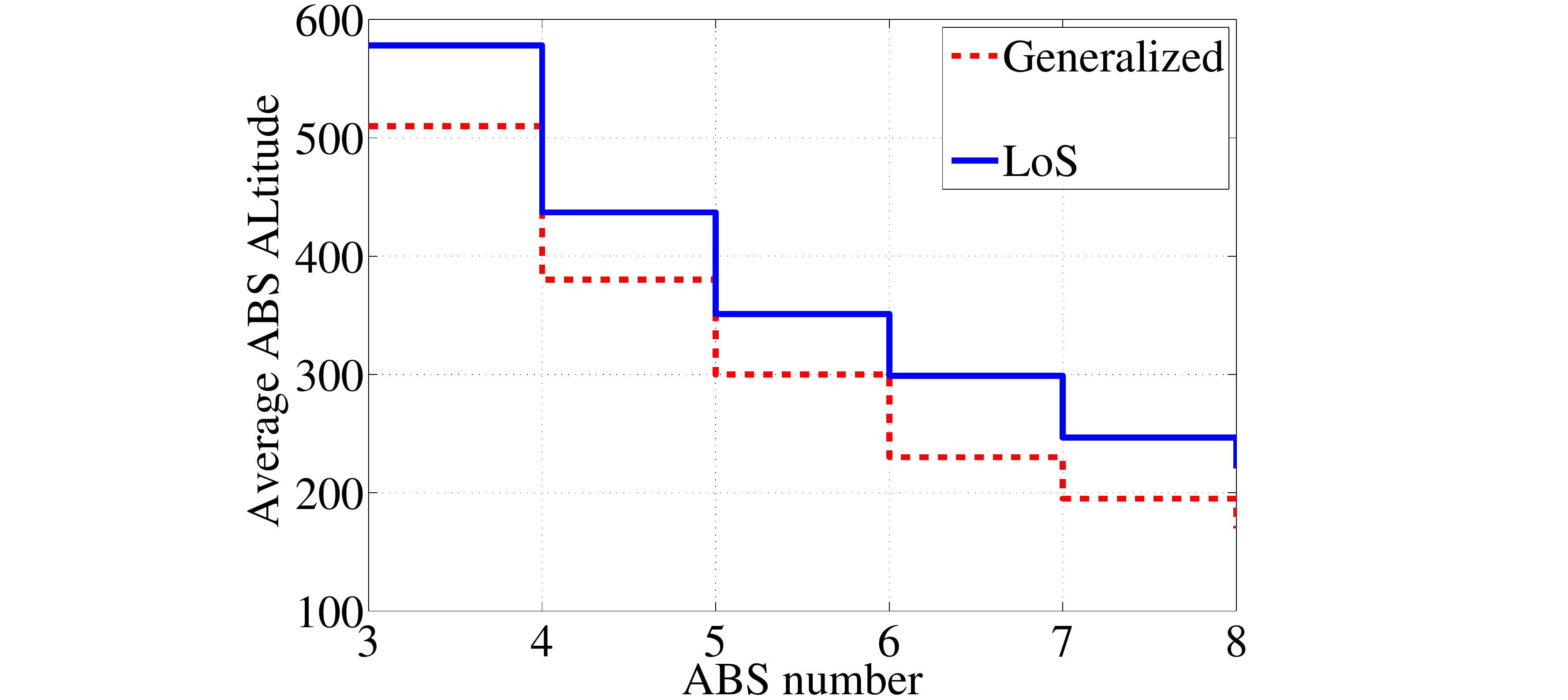}
\caption{  The average ABS altitude vs. the number of the ABSs in two cases: LoS and generalized }\label{}
\end{figure}

\begin{figure}[!h]
\hspace{0.2cm}
\includegraphics[scale=0.3]{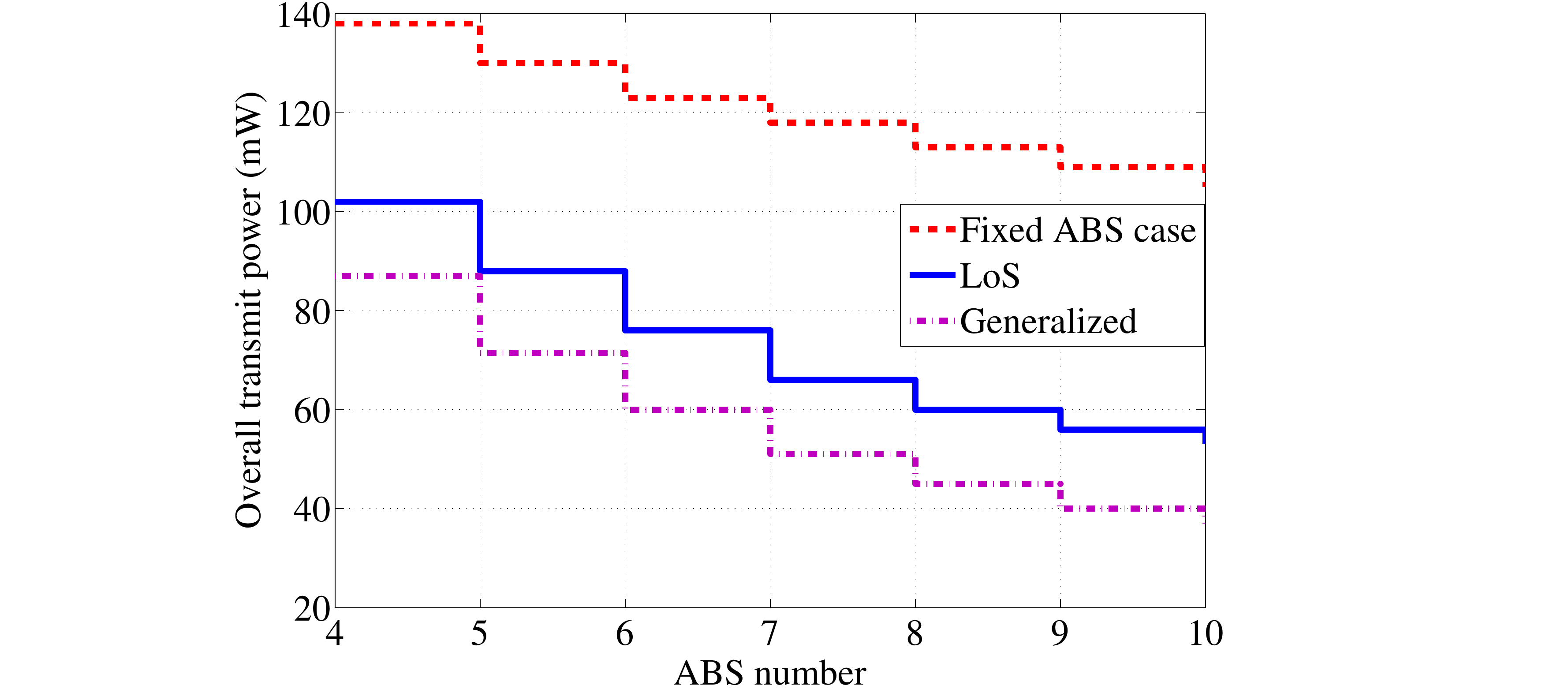}
\caption{The overall transmit power of IoT users vs. the number of the ABSs in three cases: fixed ABS case, LoS and generalized}\label{}
\end{figure}

\begin{figure}[!h]
\hspace{0.2cm}
\includegraphics[scale=0.3]{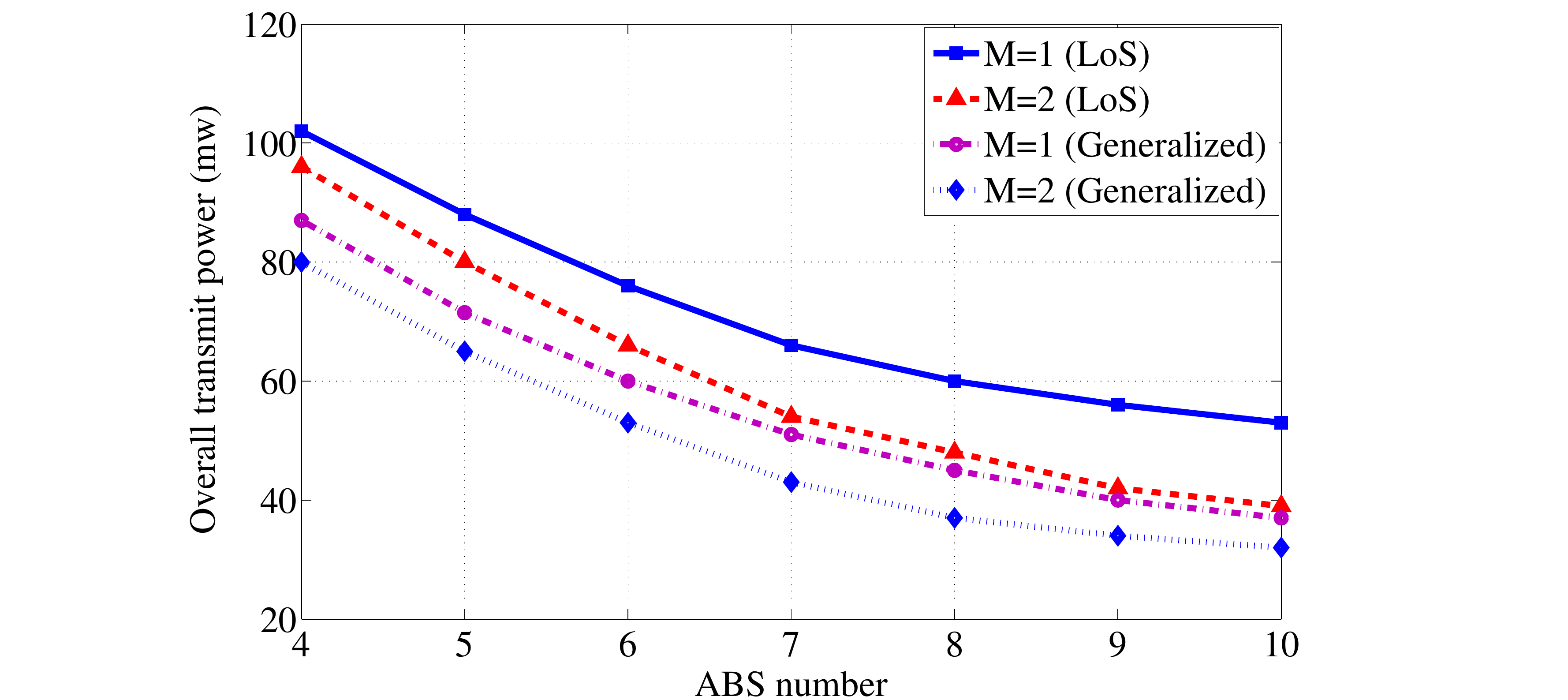}
\caption{The total transmit power of IoT users vs. the number of the ABSs in two schemes considering different modulation orders  }\label{}
\end{figure}

%\begin{figure}[!h]
%\hspace{-0.5cm}
%\includegraphics[scale=0.3]{m2tau180tau300L16}
%\includegraphics[width=8cm,height=6cm]{m2tau180tau300L16}
%\caption{ total transmit power vs. number of ABSs}\label{}
%\end{figure}
\section{Conclusion}\label{conclusion}
In this paper, we investigated the joint RRA, 3DP, and user association for ABSs in IoT networks considering adaptive modulation. Note that, we considered two schemes with different channel models, namely, generalized scheme and LoS scheme.
 In our proposed methods, we found the efficient 3D locations of ABSs in order to minimize the overal transmit power of the IoT users satisfying some existing QoS Constraint.
 The
results showed that by carefully clustering the IoT users and deploying the ABSs, the overal transmit power of the IoT users significantly 
decreases in comparison with the fixed ABSs case.
 More over, the generalized scheme can be more efficient in some practical cases in comparison with the LoS scheme. As expected, by increasing the number of ABSs, overal transmit power of users decreases. Finally, we showed that by increasing the ABSs number, the average ABS altitude decreases, and hence, it is obvious that there is a trade off between number of ABSs and average ABS altitude. Note that, we use OFDMA technology for our proposed scenario. Accordingly, we want to study on some existing multiple access technologies in next generation such as sparse code multiple access (SCMA) and power domain non orthogonal multiple access (PD-NOMA) in order to use in our proposed scenario as future work. More over, we can extend our proposed system model into multi tier heterogeneouse Base Stations which can make our model valid for next generation scenarios.
 \vspace{-0.5cm}
 \bibliography{IEEEabrv,Bibliography}

\begin{thebibliography}{9}
	\bibitem{mozaffari2016unmanned} 
	M. Mozaffari, W. Saad,  M. Bennis,  and M. Debbah, "Unmanned aerial vehicle with underlaid device-to-device communications: Performance and tradeoffs,"
	\textit{IEEE Transactions on Wireless Communications}, 
	vol. 15, no. 6, pp. 3949-3963, 2016.
	
	\bibitem{bekmezci2013flying} 
	I. Bekmezci, O. K. Sahingoz, and ¸S. Temel, "Flying ad-hoc networks (FANETs): A survey,"
	\textit{Ad Hoc Networks}, vol. 11, no. 3, pp. 1254-1270, 2013.
	
	\bibitem{orfanus2016self} 
	D. Orfanus, E. P. de Freitas, and F. Eliassen, "Self-organization as a supporting paradigm for military UAV relay networks,"
	\textit{IEEE Communications Letters}, vol. 20, no. 4, pp. 804-807, 2016.
	
	\bibitem{al2014modeling} 
	A. Hourani, S. Kandeepan, and A. Jamalipour, "Modeling air-to-ground path loss for low altitude platforms in urban environments,"
	\textit{ in Proc. of Global Communications Conference (GLOBECOM)}, Austin, TX, USA, 8-12 Dec. 2014, pp. 2898-2904.
	
	 \bibitem{motlagh2016low} 
	N. H. Motlagh, T. Taleb, and O. Arouk, "Low-altitude unmanned aerial vehicles-based internet of things services: Comprehensive survey and future perspectives,"
	\textit{IEEE Internet of Things Journal}, vol. 3, no. 6, pp. 899-922, 2016.
	
	\bibitem{razzaque2016middleware} 
	M. Razzaque, M. Milojevic, A. Palade and S. Clarke, "Middleware for internet of things: a survey,"
	\textit{IEEE Internet of Things Journal}, vol. 3, no. 1, pp. 70-95, 2016.
	
	\bibitem{angelakis2016allocation} 
	V. Angelakis, I. Avgouleas, N. Pappas, E. Fitzgerald and D. Yuan, "Allocation of heterogeneous resources of an IoT device to flexible services,"
	\textit{IEEE Internet of Things Journal}, vol. 3, no. 5, pp. 691-700, 2016.
	
	\bibitem{ngu2017iot} 
	A. Ngu, M. Gutierrez, V. Metsis, S. Nepal and Q. Sheng, "IoT middleware: A survey on issues and enabling technologies,"
	\textit{IEEE Internet of Things Journal}, vol. 4, no. 1, pp. 1-20, 2017.
	
	\bibitem{yin2017offline} 
	C. Yin, Z. Xiao, X. Cao, X. Xi, P. Yang and D. Wu, "Offline and Online Search: UAV Multi-Objective Path Planning under Dynamic Urban Environment,"
	\textit{IEEE Internet of Things Journal}, 2017.	
	  
	\bibitem{motlagh2017uav} 
	N. Motlagh, M. Bagaa and T. Taleb, "UAV-based iot platform: A crowd surveillance use case,"
	\textit{IEEE Communications Magazine}, vol. 55, no. 2, pp. 128-134, 2017.
	
	
		\bibitem{dawy2017toward} 
	Z. Dawy, W. Saad, A. Ghosh, J. G. Andrews, and E. Yaacoub, "Toward massive machine type cellular communications,"
	\textit{IEEE Wireless Communications}, 
	vol. 24, no. 1 ,pp. 120-128, 2017.
	
	\bibitem{lien2011toward} 
	S.Y. Lien, K.C. Chen, and Y. Lin, "Toward ubiquitous massive accesses in 3GPP machine-to-machine communications,"
	\textit{IEEE Communications Magazine}, 
	vol. 49, no. 4, 2011.
	
	\bibitem{dhillon2017wide} 
	H. S. Dhillon, H. Huang, and H. Viswanathan, "Wide-area wireless communication challenges for the Internet of Things,"
	\textit{IEEE Communications Magazine}, 
	vol. 55, no. 2,	pp. 168--174, 2017.
	
	
	
	\bibitem{bor2016efficient} 
	R. Yaliniz, A. El-Keyi, and H. Yanikomeroglu, "Efficient 3-D placement of an aerial base station in next generation cellular networks,"
	\textit{in Proc. of Communications (ICC)}, Kuala Lumpur, Malaysia, 22-27 May 2016, pp. 1-5.
	
		\bibitem{al2014optimal} 
A. Hourani, K. Sithamparanathan, and S. Lardner, "Optimal LAP altitude for maximum coverage,"
\textit{IEEE Wireless Communications Letters}, 
vol. 3, no. 6, pp. 569-572, 2014.
	
\bibitem{kosmerl2014base} 
J. Kosmerl and A. Vilhar, "Base stations placement optimization in wireless networks for emergency communications,"
\textit{in Proc. of Communications (ICC)}, Sydney, NSW, Australia, 10-14 June 2014, pp. 200-205.	

 \bibitem{hayajneh2016optimal} 
A. M. Hayajneh et al, "Optimal Dimensioning and Performance Analysis of Drone-Based Wireless Communications,"
\textit{in Proc. of Globecom Workshops (GC Wkshps)}, Washington, DC, USA, 4-8 Dec. 2016, pp. 1-6.


 \bibitem{alzenad2016fso} 
M. Alzenad, M. Z. Shakir, H. Yanikomeroglu, and M. Alouini, "FSO-based vertical backhaul/fronthaul framework for 5G+ wireless networks,"
\textit{arXiv preprint arXiv:1607.01472}, 2016.

	\bibitem{mozaffari2015drone} 
M. Mozaffari, W. Saad, M. Bennis, and M. Debbah, "Drone small cells in the clouds: Design, deployment and performance analysis,"
\textit{in Proc. of Global Communications Conference (GLOBECOM)},  San Diego, CA, USA, 6-10 Dec. 2015, pp. 1-6.

 \bibitem{mozaffari2016efficient} 
M. Mozaffari, W. Saad, M. Bennis, and M. Debbah, "Efficient deployment of multiple unmanned aerial vehicles for optimal wireless coverage,"
\textit{IEEE Communications Letters}, vol. 20, no. 8,  pp. 1647-1650, 2016.

 \bibitem{sharma2016uavs} 
V. Sharma, R. Sabatini and S. Ramasamy, "UAVs assisted delay optimization in heterogeneous wireless networks,"
\textit{IEEE Communications Letters}, vol. 20, no. 12,  pp. 2526-2529, 2016.

 \bibitem{sharma2016uav} 
V. Sharma, M. Bennis and R. Kumar, "UAV-assisted heterogeneous networks for capacity enhancement,"
\textit{IEEE Communications Letters}, vol. 20, no. 6,  pp. 1207-1210, 2016.

 \bibitem{merwaday2016improved} 
A. Merwaday, A. Tuncer, A. Kumbhar and I. Guvenc, "Improved throughput coverage in natural disasters: Unmanned aerial base stations for public-safety communications,"
\textit{IEEE Vehicular Technology Magazine}, vol. 11, no. 4,  pp. 53-60, 2016.

	
% \bibitem{chandrasekharan2016performance} 
%S. Chandrasekharan, A. Al-Hourani, K. Gomez,  S. Kandeepan, R. Evans, L. Reynaud, S. Scalise, "Performance evaluation of LTE and WiFi technologies in aerial networks,"
%\textit{Globecom Workshops (GC Wkshps)}, pp. 1-7, 2016.

 \bibitem{merwaday2015uav} 
A. Merwaday and I. Guvenc, "UAV assisted heterogeneous networks for public safety communications,"
\textit{in Proc. of Wireless Communications and Networking Conference Workshops (WCNCW)}, New Orleans, LA, USA, 9-12 March 2015, pp. 329-334.

 \bibitem{wang2017improving} 
Q. Wang, Z. Chen, W. Mei, J. Fang, "Improving physical layer security using UAV-enabled mobile relaying,"
\textit{IEEE Wireless Communications Letters}, 2017.
	
\bibitem{soorki2016resource} 
M. Soorki, M. Mozaffari, W. Saad, M. Manshaei and H. Saidi, "Resource allocation for machine-to-machine communications with unmanned aerial vehicles,"
\textit{in Proc. of Globecom Workshops (GC Wkshps)}, Washington, DC, USA, 4-8 Dec. 2016, pp. 1-6.

 \bibitem{li2017optimal} 
J. Li and Y. Han, "Optimal resource allocation for packet delay minimization in multi-layer UAV networks,"
\textit{IEEE Communications Letters}, vol. 21, no. 3,  pp. 580-583, 2017.
	
	\bibitem{han2009optimization} 
	Z. Han, A. L. Swindlehurst, and K. Liu, "Optimization of MANET connectivity via smart deployment/movement of unmanned air vehicles,"
	\textit{IEEE Transactions on Vehicular Technology}, 
	vol. 58, no. 7, pp. 3533-3546, 2009.
	
 \bibitem{mozaffari2016optimal} 
M. Mozaffari, W. Saad, M. Bennis, and M. Debbah, "Optimal transport theory for power-efficient deployment of unmanned aerial vehicles,"
\textit{in Proc. of Communications (ICC)}, Kuala Lumpur, Malaysia,  22-27 May 2016, pp. 1-6.
	
\bibitem{fotouhi2016dynamic} 
A. Fotouhi, M. Ding, and M. Hassan, "Dynamic base station repositioning to improve performance of drone small cells,"
\textit{in Proc. of Globecom Workshops (GC Wkshps)}, Washington, DC, USA, 4-8 Dec. 2016, pp. 1-6.

\bibitem{kalantari2017backhaul} 
E. Kalantari, M. Z. Shakir, H. Yanikomeroglu, and A. Yongac¸oglu, "Backhaul-aware robust 3D drone placement in 5G+ wireless networks,"
\textit{arXiv preprint arXiv:1702.08395}, 2017.

\bibitem{mozaffari2016mobile} 
M. Mozaffari, W. Saad, M. Bennis, and M. Debbah, "Mobile Internet of Things: Can UAVs provide an energy-efficient mobile architecture?,"
\textit{in Proc. of Global Communications Conference (GLOBECOM)}, Washington, DC, USA, 4-8 Dec. 2016
, pp. 1-6.

%\bibitem{bor2016new} 
%I. Bor-Yaliniz and H. Yanikomeroglu, "The new frontier in RAN heterogeneity: Multi-tier drone-cells,"
%\textit{IEEE Communications Magazine}, vol. 54, no.11, pp.48-55, 2016.

\bibitem{chen2017caching} 
M. Chen and M. Mozaffari, W. Saad, C. Yin, M. Debbah, C. Hong "Caching in the sky: Proactive deployment of cache-enabled unmanned aerial vehicles for optimized quality-of-experience,"
\textit{IEEE Journal on Selected Areas in Communications}, vol. 35, no.5, pp. 1046-1061, 2017.

\bibitem{li2015drone} 
X. Li, D. Guo, and H. Yin, G. Wei "Drone-assisted public safety wireless broadband network,"
\textit{in Proc. of Wireless Communications and Networking Conference Workshops (WCNCW)}, New Orleans, LA, USA, 9-12 March 2015, pp. 323-328.

\bibitem{zhan2011wireless} 
P. Zhan, K. Yu and L. Swindlehurst "Wireless relay communications with unmanned aerial vehicles: Performance and optimization,"
\textit{IEEE Transactions on Aerospace and Electronic Systems}, vol. 47, no.3, pp. 2068-2085, 2011.

\bibitem{abdulla2015toward} 
A. Abdulla, Z. Fadlullah, H. Nishiyama, N. Kato, F. Ono, R. Miura "Toward fair maximization of energy efficiency in multiple UAS-aided networks: A game-theoretic methodology,"
\textit{IEEE Transactions on Wireless Communications}, vol. 14, no. 1, pp. 305-316, 2015.

\bibitem{goldsmith2005wireless} 
 A. Goldsmith, Wireless Communications,  
Cambridge university press,
2005.

\bibitem{boyd2007tutorial} 
S. Boyd, S. Kim, L. Vandenberghe and A. Hassibi, "A tutorial on geometric programming,"
\textit{Optimization and engineering, Springer}, vol. 8, no. 1, 2007.

\bibitem{boyd2004convex} 
S. Boyd and L. Vandenberghe, "A tutorial on geometric programming,"
\textit{Cambridge university press}, 2004.

\bibitem{luo2010semidefinite} 
Z. Luo, W. Ma, A. So, Y. Ye and S. Zhang "Semidefinite relaxation of quadratic optimization problems,"
\textit{IEEE Signal Processing Magazine}, vol. 27, no. 3, pp. 20-34, 2010.
%		\bibitem{pang2014efficient} 
%	Y. Pang, Y. Zhang, Y. Gu, M. Pan, Z. Han, and P. Li, "Efficient data collection for wireless rechargeable sensor clusters in harsh terrains Using UAVs,"
%	\textit{Global Communications Conference (GLOBECOM)}, 
% pp. 234-239, 2014.

%   	\bibitem{abuzainab2016cognitive} 
% N. Abuzainab, W. Saad, and H. V. Poor, "Cognitive hierarchy theory for heterogeneous uplink multiple access in the Internet of Things,"
% \textit{Information Theory (ISIT)}, 
 %pp. 1252-1256, 2016.
 
 
	
%	\bibitem{tu2011energy} 
%	C.Y. Tu, C.Y. Ho, and C.Y. Huang, "Energy-efficient algorithms and evaluations for massive access management in cellular based machine to machine communications,"
%	\textit{Vehicular Technology Conference (VTC)}, 
%	
	
%		\bibitem{bucaille2013rapidly} 
%	I. Bucaille, S. Hethuin, A. Munari, R. Hermenier, T. Rasheed, and
%	S. Allsopp, "Rapidly deployable network for tactical applications: Aerial base station with opportunistic links for unattended and temporary events absolute example,"
%	\textit{Military Communications Conference}, 
%	pp. 1116-1120, 2013.
	
	

%\bibitem{feng2006wlcp2} 
%Q. Feng, E. K. Tameh, A. R. Nix, and J. McGeehan, "Modelling the likelihood of line-of-sight for air-to-ground radio propagation in urban environments,"
%\textit{Global Telecommunications Conference (GLOBECOM)}, 	pp. 1-5, 2006.

%	\bibitem{feng2006path} 
%Q. Feng, J. McGeehan, E. K. Tameh, and A. R. Nix, "Path loss models for air-to-ground radio channels in urban environments,"
%\textit{Vehicular Technology Conference (VTC)}, 
%vol. 6, pp. 2901-2905, 2006.

%	\bibitem{holis2008elevation} 
%J. Holis and P. Pechac, "Elevation dependent shadowing model for mobile communications via high altitude platforms in built-up areas,"
%\textit{IEEE Transactions on Antennas and Propagation}, 
%vol. 56, no. 4, pp. 1078-1084, 2008.	
	 
%	 \bibitem{daniel2011using} 
%	 K. Daniel and C. Wietfeld, "Using public network infrastructures for UAV remote sensing in civilian security operations,"
%	 \textit{DORTMUND UNIV (GERMANY FR)}, 
%	 pp. 200-205, 2014.
	 
%	 \bibitem{rohde2012interference} 
%	 S. Rohde and C. Wietfeld, "Interference aware positioning of aerial relays for cell overload and outage compensation,"
%	 \textit{Vehicular Technology Conference (VTC)}, 
%	 pp. 1-5, 2012.
	 
%	 \bibitem{jiang2012optimization} 
%	 F. Jiang and A. L. Swindlehurst, "Optimization of UAV heading for the ground-to-air uplink,"
%	 \textit{IEEE Journal on Selected Areas in Communications (JSAC)}, vol. 30, no. 5, pp. 993-1005, 2012.

	 
	 
%	  \bibitem{yaacoub2012energy} 
%	 E. Yaacoub and O. Kubbar, "Energy-efficient device-to-device communications in LTE public safety networks,"
%	 \textit{Globecom Workshops (GC Wkshps)}, pp. 391-395, 2012.
	 
%	  \bibitem{doppler2009device} 
%	 K. Doppler, M. Rinne, C. Wijting, C. B. Ribeiro, and K. Hugl, "Device-to-device communication as an underlay to LTE-advanced networks,"
%	 \textit{IEEE Communications Magazine}, vol. 47, no. 12, pp. 42-49, 2009.
	 
%	 \bibitem{lee2015power} 
%	N. Lee, X. Lin, J. G. Andrews, and R. Heath, "Power control for D2D underlaid cellular networks: Modeling, algorithms, and analysis,"
%	 \textit{IEEE Journal on Selected Areas in Communications (JSAC)}, vol. 33, no. 1, pp. 1-13, 2015.
	 
%	 \bibitem{shalmashi2016energy} 
%	S. Shalmashi, E. Bj¨ornson, M. Kountouris, K. W. Sung, and M. Debbah, "Energy efficiency and sum rate tradeoffs for massive MIMO systems with underlaid device-to-device communications,"
%	 \textit{EURASIP Journal on Wireless Communications and Networking}, p. 175, 2016.
	 
%	 \bibitem{lin2015interplay} 
%	 X. Lin, R. Heath, and J. Andrews, "The interplay between massive MIMO and underlaid D2D networking,"
%	 \textit{IEEE Transactions on Wireless Communications}, vol. 14, no. 6, pp. 3337-3351, 2015.
	  
	 
%	  \bibitem{sharma2017intelligent} 
%	 V. Sharma, K. Srinivasan, H.-C. Chao, K.-L. Hua, and W.-H. Cheng, "Intelligent deployment of UAVs in 5G heterogeneous communication environment for improved coverage,"
%	 \textit{Journal of Network and Computer Applications}, vol. 85, pp. 94-105, 2017.
	 
%	  \bibitem{de2013cooperation} 
%	 E. P. de Freitas, T. Heimfarth, A. Vinel, F. R. Wagner,
%	 C. E. Pereira, and T. Larsson, "Cooperation among wirelessly connected static and mobile sensor nodes for surveillance applications,"
%	 \textit{Sensors}, vol. 13, no. 10, pp. 12903-12928, 2013.
	 
	 
%	  \bibitem{kandeepan2014aerial} 
%	 S. Kandeepan, K. Gomez, L. Reynaud, and T. Rasheed, "Aerial-terrestrial communications: terrestrial cooperation and energy-efficient transmissions to aerial base stations,"
%	 \textit{IEEE Transactions on Aerospace and Electronic Systems}, vol. 50, no. 4, pp. 2715-2735, 2014.
	 
	  
%	 \bibitem{atkins2010risk} 
%	 E. M. Atkins, "Risk identification and management for safe UAS operation,"
%	 \textit{Systems and Control in Aeronautics and Astronautics (ISSCAA)}, pp. 774-779, 2010.
	 
	 
%	  \bibitem{sathyanarayanan2016designing} 
%	S. Chandrasekharan, K. Gomez, A. Al-Hourani, S. Kandeepan,
%	T. Rasheed, L. Goratti, L. Reynaud, D. Grace, I. Bucaille, T. Wirth, and
%	S. Allsopp
%	, "Designing and Implementing Future Aerial Communication Networks,"
%	 \textit{IEEE communications magazine}, vol. 54, no.5, pp.26-34, 2016.




	
	 
\end{thebibliography}

\end{document}